%% file: 00-main.tex
\documentclass[times]{mcp-acm}

\input{idp-latex/idp-latex}
\usepackage{graphicx}
\usepackage{tikz}
\usetikzlibrary{shapes,arrows,snakes,positioning,calc}
\newcommand{\il}[1]{\lstinline{#1}\xspace}
\newcommand{\inputstar}{input\text{\textasteriskcentered}\xspace}
\newcommand{\outputstar}{output\text{\textasteriskcentered}\xspace}

\begin{document}
\frontmatter
\mainmatter


\chapter{Predicate Logic as a Modelling Language: \\ The \idp System}

AUTHORS: Broes De Cat, Bart Bogaerts,  Maurice Bruynooghe, 
  Gerda Janssens, Marc Denecker
  
  Department of Computer Science, KU Leuven


\input{1-abstract}
\input{1-introduction}
\input{2-preliminaries}

\input{3a-idp}

\input{3b-language}

\input{4-mx}

\input{5-modeling}

\input{6-related}

\section{Conclusion}
Kowalski's 1974 paper \cite{ifip/Kowalski74} laid the foundations for the field of Logic Programming, by giving the Horn-clause subset of predicate logic a procedural interpretation to use it for programming.
More recently, progress in automated reasoning in fields such as SAT and CP made the exploration possible of more pure forms of declarative programming, gradually moving from declarative programming to declarative modelling, in which the user only has to care about the problem specification.

In this chapter, we took this development one step further and presented the knowledge base system \idp, in which knowledge is separated from computation.
The knowledge representation language is both natural and extensible, cleanly integrating first-order logic with definitions, aggregates, etc.
It provides a range of inference engines and functionalities for tasks encountered often in practice.

\idp is an extensible framework for declarative modelling, in which both language extensions and inference engines can be added with relative ease.
It focuses on moving the burden of performance on modelling from the user to the system, demonstrated by the workflow of optimization inference, which is achieved by combining insights from fields such as SAT, constraint programming, logic programming and answer set programming.



\clearpage

\bibliography{idp-latex/krrlib}

\end{document}

%% file: idp-latex/idp-latex.tex
\usepackage{import}

\subimport{idp-latex/}{includes-no-theorems}
\subimport{idp-latex/}{common-citations}

\subimport{idp-latex/}{theorems}

\ignore{

\include{includes-no-theorems}
\include{common-citations}
\include{theorems}
}

%% file: idp-latex/includes-no-theorems.tex
\usepackage{ifthen}
\usepackage{url}
\usepackage{amssymb}
\usepackage{amsmath}
\usepackage{import}
\subimport{packages/}{xparse}
\usepackage{amsmath}
\usepackage[usenames,dvipsnames]{xcolor}
\usepackage{xspace}
\usepackage{datatool}
\usepackage{glossaries}

\providecommand\m[1]{\ensuremath{{#1}}\xspace}
\renewcommand{\m}[1]{\ensuremath{{#1}}\xspace}
\providecommand\mungrouped[1]{\ensuremath{#1}\xspace}
\renewcommand{\mungrouped}[1]{\ensuremath{#1}\xspace}
\newcommand{\trval}[1]{\m{\bf #1}}



	\newcommand{\limplies}{\Rightarrow}
	
	\newcommand{\lequiv}{\Leftrightarrow}
	\newcommand{\limpliedby}{\Leftarrow}
	
	\newcommand{\lrule}{\leftarrow}
	\newcommand{\cause}{\stackrel{c}{\lrule}}
	\newcommand{\rul}{\leftarrow}
	\newcommand{\ltrue}{\trval{t}}
	\newcommand{\lfalse}{\trval{f}}
	\newcommand{\lunkn}{\trval{u}}
	
	\newcommand{\bigand}{\bigwedge}
	\newcommand{\bigor}{\bigvee}


	\newcommand{\voc}{\m{\Sigma}}

	\newcommand{\struct}{\m{I}}
	
	\newcommand{\I}{\m{\mathcal{I}}}
	
	\newcommand{\J}{\m{\mathcal{J}}}

	\newcommand{\theory}{\m{\mathcal{T}}}


	\newcommand{\D}{\m{\Delta}}
	\newcommand{\f}{\m{\varphi}}

	\NewDocumentCommand\inter{g+g}{%
	  \IfNoValueTF{#1}
	    {\struct}
	    {\m{#1^{#2}}}}



	\newcommand{\xxx}{\m{\overline{x}}}
	
	\newcommand{\yyy}{\m{\overline{y}}}
	
	\newcommand{\ddd}{\m{\overline{d}}}
	
	\newcommand{\ccc}{\m{\overline{c}}}

	\newcommand{\ttt}{\m{\overline{t}}}

 	
	\newcommand{\TTT}{\m{\overline{T}}}



	\newcommand{\bool}{\m{\mathbb{B}}}
	
	\newcommand{\nat}{\m{\mathbb{N}}}
	
	\renewcommand{\int}{\m{\mathbb{Z}}}

	\newcommand{\leqp}{{\m{\,\leq_p\,}}}




	\NewDocumentCommand\subs{g+g}{%
	  \IfNoValueTF{#1}
	    {\m{/}}
	    {\m{#1/ #2}}}


	\newcommand{\logicname}[1]{\textsc{#1}\xspace}

	\newcommand{\idp}{\logicname{IDP}}
	\newcommand{\xsb}{\logicname{XSB}}
	\newcommand{\idptwo}{\logicname{IDP$^2$}}
	\newcommand{\idpthree}{\logicname{IDP3}}

	\newcommand{\minisat}{\logicname{MiniSAT}}
	\newcommand{\minisatid}{\logicname{MiniSAT(ID)}}
	\newcommand{\gidl}{\logicname{GidL}}


	\newcommand{\foid}{\logicname{FO(ID)}}
	\newcommand{\foidaggpf}{\logicname{FO(ID,\allowbreak Agg,\allowbreak PF)}}
	\newcommand{\foidaggpft}{\logicname{FO(ID,\allowbreak Agg,\allowbreak PF,\allowbreak T)}}

\newcommand{\ouracronym}[3]{%
	\newacronym{#1}{#2}{#3}
	\expandafter\newcommand\csname #1\endcsname{\gls{#1}\xspace}%
}
	\ouracronym{FO}{FO}{first-order logic}
	\ouracronym{MX}{MX}{Model Expansion}
	\ouracronym{MO}{MO}{Model Optimization}
	\ouracronym{ASP}{ASP}{Answer Set Programming}
	\ouracronym{TP}{TP}{Theorem Proving}
	\ouracronym{LP}{LP}{Logic Programming}
	\ouracronym{CP}{CP}{Constraint Programming}
	\ouracronym{FP}{FP}{Functional Programming}
	\ouracronym{KR}{KR}{Knowledge Representation}
	\ouracronym{CSP}{CSP}{Constraint Satisfaction Problem}
	\ouracronym{SMT}{SMT}{SAT Modulo Theories}
	\ouracronym{KBS}{KBS}{knowledge base system}
	\ouracronym{NNF}{NNF}{Negation Normal Form}
	\ouracronym{FNNF}{FNNF}{Flat Negation Normal Form}
	\ouracronym{DefNNF}{DefNNF}{Definition Negation Normal Form}
	\ouracronym{CDCL}{CDCL}{Conflict-Driven Clause-Learning}
	\ouracronym{LCG}{LCG}{Lazy Clause Generation}
	\ouracronym{AEL}{AEL}{Autoepistemic Logic}
	\ouracronym{OEL}{OEL}{Ordered Epistemic Logic}
	\ouracronym{AFT}{AFT}{Approximation Fixpoint Theory}



%

	\makeatletter
	\def\ifenv#1{
	\def\@tempa{#1}%
	\def\@ttempa{#1*}%
	\ifx\@tempa\@currenvir
	\expandafter\@firstoftwo
	\else
	\expandafter\@secondoftwo
	\fi
	}
	\makeatother

	\newcommand{\ddrule}[4]{\ensuremath{#1 \leftarrow #2 & \{#3\} & #4}}
	\newcommand{\drule}[2]{\ensuremath{#1 & \leftarrow & #2}}

	\newcommand{\darule}[4]{\ensuremath{#1 \leftarrow #2 & \{#3\} & #4}}
	\newcommand{\arule}[2]{\ensuremath{#1 \, &\leftarrow \, #2}}

	\newenvironment{ldef}{\left\{\begin{array}{l@{ \,}l@{\,}l}}{\end{array}\right\}}
	\newenvironment{ltheo}{\[\begin{array}{l}}{\end{array}\]\ignorespacesafterend}

	\newcommand{\LNDRule}[2]{
	\ifenv{array}
	{\drule{#1}{#2}}
	{ \ifenv{align}
		{\arule{#1}{#2}}
		{\ifenv{align*}
		{\arule{#1}{#2}}
		{ERROR: using LDRule in unsupported environment: \@currenvir}
		}
	}
	}

	\newcommand{\LDRule}[4]{
	\ifenv{array}
	{\ddrule{#1}{#2}{#3}{#4}}
	{ \ifenv{align}
		{\darule{#1}{#2}{#3}{#4}}
		{\ifenv{align*}
		{\darule{#1}{#2}{#3}{#4}}
		{ERROR: using LDRule in unsupported environment: \@currenvir}
		}
	}
	}

	\NewDocumentCommand\LRule{m+g+g+g}{%
		\IfNoValueTF{#2}%
		{#1.&}{%
		\IfNoValueTF{#3}
		{\LNDRule{#1}{#2.}}
		{\LDRule{#1}{#2.}{#3}{#4}}%
		}
	}


	\NewDocumentCommand\CLRule{m+g}{%
	\ifenv{array}
	{\cdrule{#1}{#2}}
	{ \ifenv{align}
		{\carule{#1}{#2}}
		{\ifenv{align*}
			{\carule{#1}{#2}}
			{ERROR: using CLRule in unsupported environment: \@currenvir}
		}
	}
	}

	\NewDocumentCommand\carule{m+g}{%
		\IfNoValueTF{#2}
			{\ensuremath{#1.}}
			{\ensuremath{#1 \, &\cause \, #2}}}
	\NewDocumentCommand\cdrule{m+g}{%
		\IfNoValueTF{#2}
			{\ensuremath{#1.}}
			{\ensuremath{#1 & \cause & #2}}}



	\newcommand{\algrule}[4]{
	\hbox{{#1}:}& 
	\quad #2 ~\longrightarrow~ #3 
	\hbox{~ if } #4\\
	}

	\newcommand{\AlgoRule}[4]{
	\ifenv{array}
	{\algrule{#1}{#2}{#3}{#4}}
		{ERROR: using AlgoRule in unsupported environment: \@currenvir}
	}

\newcommand{\commentstyle}{\color{Gray}}

	\usepackage[final]{listings}

	\lstdefinelanguage{idp}{
		morekeywords=[1]{query(}, 
		morekeywords=[2]{namespace,vocabulary,theory,structure,procedure,term,set,formula, spec, specification,query},
		morekeywords=[3]{include,using,type,isa,contains,partial,extern,LFD,GFD,constructed,from,constraint,pred,supertype,of,subtype,define},
		morekeywords=[4]{int,float,char,string,nat},
		morekeywords=[5]{if,then,else,for,end},
		morecomment=[s]{/*}{*/},	
		morecomment=[l]{//}
	}
	\lstset{
		language=idp,
		tabsize=3,
		basicstyle=\small,
		frame=none,
		frame=single,
		showstringspaces=false,
		breaklines = true,
		alsoletter=(,
		commentstyle=\commentstyle,
		keywordstyle=[1],
		keywordstyle=[2]\color{BrickRed}\bfseries,
		keywordstyle=[3]\color{OliveGreen}\bfseries,
		keywordstyle=[4]\color{Blue}\bfseries,
		keywordstyle=[5]\color{Violet}\bfseries,
		literate={~} {$\sim$}{1}
	}


	\newcommand{\ignore}[1]{}

	\newboolean{nocomments}
	\setboolean{nocomments}{false}

	\newboolean{commentmargin}
	\setboolean{commentmargin}{true}

	\newcommand{\namedcomment}[3]{
		\ifthenelse{\boolean{nocomments}}
		{} 
		{ 
			\ifthenelse{\boolean{commentmargin}}
				{ {\color{#3} \marginpar{\color{#3}\sc #2}#1}  } 
				{  {\color{#3} {\sc #2}: #1}  } 
		}
	}
	\newcommand{\mnamedcomment}[3]{\ifthenelse{\boolean{nocomments}}{}{{\marginpar{ \color{#3}{\sc #2}:#1}}}}


	\newcommand{\bart}[1]{\namedcomment{#1}{bb}{red}}

	\newcommand{\marc}[1]{\namedcomment{#1}{md}{orange}}

	\newcommand{\maurice}[1]{\namedcomment{#1}{mb}{orange}}

	\usepackage{soul}




%% file: idp-latex/packages/xparse.tex
\usepackage{xparse}

%% file: idp-latex/common-citations.tex
\usepackage{etoolbox}

\newcommand\setcitation[2]{%
  \csdef{mycommoncitation#1}{#2}}
\newcommand\getcitation[1]{%
  \csuse{mycommoncitation#1}}

\setcitation{IDP}{WarrenBook/DeCatBBD14}
\setcitation{idp}{WarrenBook/DeCatBBD14}
\setcitation{fodot}{tocl/DeneckerT08}
\setcitation{CP}{Apt03}
\setcitation{cp}{Apt03}
\setcitation{EZCSP}{lpnmr/Balduccini11}
\setcitation{KR}{Baral:2003}
\setcitation{ASPComp2}{lpnmr/DeneckerVBGT09}
\setcitation{ASPComp3}{journals/tplp/CalimeriIR14}
\setcitation{ASPComp4}{conf/lpnmr/AlvianoCCDDIKKOPPRRSSSWX13}
\setcitation{CPSupport}{ictai/DeCat13}
\setcitation{CPsupport}{ictai/DeCat13}
\setcitation{functionDetection}{iclp/DeCatB13}
\setcitation{fodot2asp}{DeneckerLTV12} 
\setcitation{Tarskian}{DeneckerLTV12} 
\setcitation{Inca}{iclp/DrescherW12}
\setcitation{csp2asp}{ijcai/DrescherW11}
\setcitation{DPLLT}{cav/GanzingerHNOT04}
\setcitation{AspInPractice}{synthesis/2012Gebser}
\setcitation{clasp}{ai/GebserKS12}
\setcitation{oclingo}{kr/GebserGKOSS12}
\setcitation{gringo}{lpnmr/GebserST07}
\setcitation{cmodels}{aaai/GiunchigliaLM04}
\setcitation{inputster}{tplp/Jansen13}
\setcitation{DLV}{tocl/LeonePFEGPS06}
\setcitation{LearningPaper}{TPLP/BruynoogheBBDDJLRDV} 
\setcitation{clog}{iclp/BogaertsVDV14} 
\setcitation{foc}{iclp/BogaertsVDV14}
\setcitation{FOC}{iclp/BogaertsVDV14}
\setcitation{inferenceClog}{ecai/BogaertsVDV14}
\setcitation{examplesClog}{nmr/BogaertsVDV14b} 
\setcitation{AFT}{DeneckerBV12}
\setcitation{KBS}{iclp/DeneckerV08}
\setcitation{KBS-invitedtalk}{jelia/Denecker16}
\setcitation{KBPE}{inap/DePooterWD11}
\setcitation{lazyGrounding}{jair/CatDBS15} 
\setcitation{lazygrounding}{jair/CatDBS15} 
\setcitation{ASP}{marek99stable}
\setcitation{satid}{sat/MarienWDB08}
\setcitation{lazyclausegeneration}{constraints/OhrimenkoSC09}
\setcitation{FP}{ACMCS/Hudak89}
\setcitation{GroundingWithBounds}{jair/WittocxMD10}
\setcitation{SAT}{faia/SilvaLM09}
\setcitation{LTC}{iclp/Bogaerts14}
\setcitation{SPSAT}{ictai/DevriendtBMDD12}
\setcitation{BreakID}{sat/DevriendtBBD16}
\setcitation{breakid}{sat/DevriendtBBD16}
\setcitation{LCG}{stuckeyLCG}
\setcitation{MiniZinc}{conf/cp/NethercoteSBBDT07}
\setcitation{minizinc}{conf/cp/NethercoteSBBDT07}
\setcitation{amadini}{cpaior/AmadiniGM13}
\setcitation{bootstrapping}{ngc/BogaertsJDJBD16}
\setcitation{Bootstrapping}{ngc/BogaertsJDJBD16}
\setcitation{GroundedFixpoints}{ai/BogaertsVD15}
\setcitation{PartialGroundedFixpoints}{ijcai/BogaertsVD15}
\setcitation{LogicBlox}{datalog/GreenAK12}
\setcitation{proB}{journals/sttt/LeuschelB08}
\setcitation{NaturalInductions}{KR/DeneckerV14} 
\setcitation{LP}{jacm/EmdenK76}
\setcitation{SMT}{faia/BarrettSST09}
\setcitation{AF}{ai/Dung95}
\setcitation{ADF}{kr/BrewkaW10}
\setcitation{af}{ai/Dung95}
\setcitation{adf}{kr/BrewkaW10}
\setcitation{ADFRevisited}{ijcai/BrewkaSEWW13}
\setcitation{adfrevisited}{ijcai/BrewkaSEWW13}
\setcitation{DefaultLogic}{ai/Reiter80}
\setcitation{DL}{ai/Reiter80}
\setcitation{AEL}{mo85}
\setcitation{minisat}{sat/EenS03}
\setcitation{completion}{adbt/Clark78}
\setcitation{wasp}{lpnmr/AlvianoDFLR13}
\setcitation{minisatid}{ictai/DeCat13}
\setcitation{lcg}{stuckeyLCG}
\setcitation{CEGAR}{jacm/ClarkeGJLV03}
\setcitation{cegar}{jacm/ClarkeGJLV03}
\setcitation{CuttingPlane}{or/DantzigFJ54}
\setcitation{kodkod}{tacas/TorlakJ07}
\setcitation{cdcl}{Marques-SilvaS99}
\setcitation{CDCL}{Marques-SilvaS99}
\setcitation{relevance}{ijcai/JansenBDJD16}
\setcitation{relevance-implementation}{aspocp/JansenBDJD16}
\setcitation{WFS}{GelderRS91}
\setcitation{wfs}{GelderRS91}
\setcitation{UnfoundedSet}{GelderRS91}
\setcitation{stablesemantics}{iclp/GelfondL88}
\setcitation{StableSemantics}{iclp/GelfondL88}
\setcitation{shatter}{Shatter}
\setcitation{sbass}{drtiwa11a}
\setcitation{lparsemanual}{url:lparse_manual}
\setcitation{AIC}{ppdp/FlescaGZ04}
\setcitation{templates}{tplp/DassevilleHJD15}
\setcitation{templates2}{iclp/DassevilleHBJD16}
\setcitation{sat-to-sat}{aaai/JanhunenTT16}
\setcitation{sat-to-sat-qbf}{bnp/BogaertsJT16}
\setcitation{sat-to-sat-QBF}{bnp/BogaertsJT16}
\setcitation{sat-to-sat-SO}{kr/BogaertsJT16}
\setcitation{XSB}{SwiW12}
\setcitation{KCmap}{jair/DarwicheM02}
\setcitation{TLA}{DBLP:books/aw/Lamport2002}
\setcitation{EventB}{BookAbrial2010}
\setcitation{MX}{MitchellTHM06}
\setcitation{MIP}{Sierksma96}
\setcitation{perefectmodel}{minker88/Przymusinski88}
\setcitation{SafeInductions}{ijcai/BogaertsVD17}
\setcitation{AIC}{ppdp/FlescaGZ04}
\setcitation{aic}{ppdp/FlescaGZ04}
\setcitation{SLDNF}{Lloyd87}
\setcitation{}{}
\setcitation{}{}
\setcitation{}{}
\setcitation{}{}
\setcitation{}{}
\setcitation{}{}
\setcitation{}{}
\setcitation{}{}
\setcitation{}{}
\setcitation{}{}
\setcitation{}{}
\setcitation{}{}
\setcitation{}{}
\setcitation{}{}
\setcitation{}{}
\setcitation{}{}
\setcitation{}{}
\setcitation{}{}
\setcitation{}{}
  
\newcommand\refto[1]{%
      \getcitation{#1}}
      
\newcommand\mycite[1]{%
      \ifcsname mycommoncitation#1\endcsname%
   \cite{\getcitation{#1}}%
  \else%
    \cite{#1}%
  \fi%
}	
  

%% file: idp-latex/theorems.tex
\usepackage{amsthm}

\theoremstyle{plain}
\newtheorem{thm}{Theorem}[section]

\newtheorem*{lem*}{Lemma}

\theoremstyle{definition}

\newtheorem{definition}[thm]{Definition}

\newtheorem*{nota*}{Notation}

\newtheorem*{note*}{Note}

\newtheoremstyle{example_basic} 
{\topsep} 
{\topsep} 
{\normalfont}
{0pt}
{\bfseries}
{.}
{5pt plus 1pt minus 1pt}
{}

\newtheoremstyle{example_contd}
{\topsep} 
{\topsep} 
{\normalfont}
{0pt}
{\bfseries}
{.}
{5pt plus 1pt minus 1pt}
{\thmname{#1} \thmnumber{ #2}\thmnote{#3} (continued)}

\theoremstyle{example_basic} 

\newtheorem{example}[thm]{Example}

\newtheorem{ex*}{Example}
\theoremstyle{example_contd}

\theoremstyle{plain}

%% file: 1-abstract.tex
  With the technology of the time, Kowalski's seminal 1974 paper {\em
    Predicate Logic as a Programming Language} was a breakthrough for
  the use of logic in computer science. It introduced two fundamental
  ideas: on the declarative side, the use of the Horn clause logic
  fragment of classical logic, which was soon extended with negation
  as failure, on the procedural side the procedural
  interpretation 
  which made it possible to write algorithms in the formalism.

  Since then, strong progress was made both on the declarative
  understanding of the logic programming formalism and in automated
  reasoning technologies, particularly in SAT solving, Constraint
  Programming and Answer Set Programming. This has paved the way for
  the development of an extension of logic programming that embodies a
  more pure view of logic as a modelling language and its role for
  problem solving.  

  In this chapter, we present the \idp language and system. The language
  is essentially classical logic extended with one of logic
  programmings most important contributions to knowledge
  representation: the representation of complex definitions as rule
  sets under well-founded semantics. The system is a knowledge base
  system: a system in which complex declarative information is stored
  in a knowledge base which can be used to solve different
  computational problems by applying multiple forms of inference.  In
  this view, theories are declarative modellings, bags of information, i.e.,
  descriptions of possible states of affairs. They are neither
  procedures nor descriptions of computational problems. As such, the
  \idp language and system preserve the fundamental idea of a
  declarative reading of logic programs, while they break with the
  fundamental idea of the procedural interpretation of logic programs.


\ignore{ To this day, these features are viewed by many as the
  defining properties of logic programming formalisms.

  What \idp retains from logic programming is, on the declarative
  level, the use of rules to express definitions and, on the
  computational level, the goal to solve computational problems with
  logic.  What is different compared to logic programming is that
  theories are pure declarative modellings; they do not have a
  procedural interpretation, in fact they are not even representations
  of computional problems. They are bags of information, descriptions
  of possible states of affairs.

  One of its main language constructs is the definition, which
  directly From logic programming, it retains the use of rules, not as
  procedures but as representations of definitional knowledge.

  In this view, theories are descriptions of states of affairs, not of
  procedures or even (computational) problems. This is a major break
  with a fundamental idea of logic programming. But while

The more recent    tremendous progress in automated reasoning technologies,
    particularly in SAT solving, Constraint Programming has paved the way for the use of
    logic as a modelling language.

This chapter describes the realisation of such a modelling language as the \idp \KBS.  
In contrast to declarative programming, the user only specifies her
knowledge about a problem and does not need to pay attention to control
issues. In the \idp system, declarative modelling is done in the
\idp language which combines inductive definitions (similar to
sets of Prolog rules) with first-order logic, types and aggregates;
this facilitates concise specifications.

This chapter presents the language, motivates the design choices and
gives an overview of the system architecture and the implementation
techniques. It also gives an overview of different inference methods
supported by the system such as query evaluation, model expansion and
theorem proving, and explains in detail how combining various functionalities results in a state-of-the-art model expansion engine.
Finally, it explains how a tight integration with a procedural language (Lua) allows users to treat logical components as first-class citizens and to solve complex problems in a workflow of (multi-inference) interactions.
\end{abstract}

}

%% file: 1-introduction.tex
\section{Introduction}  

Since the early days of artificial intelligence, it is believed that
logic could bring important benefits in solving computational problems
and tasks compared to standard programming languages.  Kowalski's
seminal paper {\em Predicate Logic as a Programming
  Language}~\cite{ifip/Kowalski74} was a major step in this direction
and laid the foundations for the field of logic programming. It introduced two
fundamental ideas: on the declarative level, the use of the Horn
clause logic fragment of classical logic; on the procedural level, a 
procedural interpretation of this logic which made it possible to
write algorithms in the formalism. With the technology of the time,
Kowalski's paper was a breakthrough for the use of logic in computer
science.

Since then, logic programming has fanned out in many directions, but
in most extensions and variants, the original key ideas are still
present in stronger or weaker form: the use of a rule-based formalism,
and the presence of a procedural interpretation.  Or at least, if
programs are not procedures, they are representations of 
computational problems, as in Datalog and in Answer Set programming.

In this chapter, we present the \idp system (and language) that, although
it builds on the accomplishments of logic programming and contains
language constructs based on logic programming, embodies a more pure
view of logic as a modelling language. In this view, a logic theory is
not a program, it cannot be executed or run; it does not describe an
algorithm. A theory, in principle, is not even a specification of a
problem. A theory is a bag of information, a description of possible
states of affairs of the domain of discourse, a
representation of knowledge, or in other words a modelling of the domain of
discourse. As such, this view breaks the link that Kowalski had laid
between logic and programming.  On the other hand, the \idp language
contains a language construct, namely inductive definition, that directly descends from logic
programming, and the \idp system is designed to \emph{use} declarative
information to solve problems and uses many technologies that were
developed in logic programming. As such, the \idp language and system
preserve some of the contributions of logic programming but break with some of the
fundamental ideas of logic programming.

\renewcommand{\LP}{\color{red}TODO REMOVE ME}

To explain the \idp language, we need to go back to the early days of
logic programming when negation as failure was introduced. On the one
hand, conclusions obtained with the negation as failure inference rule
were often natural and as desired. This is illustrated by the program and query
in Table~\ref{tabNAF}.
\begin{table}\begin{center}
\begin{minipage}{6cm}
\begin{align*}
&\tt Human(John).				&&\text{\tt John is a Human}\\
&\tt Human(Jane).				&&\text{\tt Jane is a Human}\\
&\tt Male(John).   &&\text{\tt John is a Male}\\
&\tt Female(x) :- Human(x), not Male(x). &&\text{\tt Females are Humans that are not Male}\\
& \tt?- Female(Jane). &&\text{\tt is Jane Female?}\\
yes
\end{align*}
\end{minipage}  \end{center}
  \caption{Prolog answers ``yes'' to the query {\tt ?- Female(Jane).} \label{tabNAF}}
\end{table}
On the other hand, these answers were logically unsound if rules were
interpreted as material implications\footnote{This is, the logical implication $\varphi\limplies \psi$ that is interpreted as $\lnot \varphi \lor \psi$.}. This problem disturbed the logic
programming community for more than a decade and led to the
development of stable and well-founded semantics \cite{iclp/GelfondL88,GelderRS91}. However, a
formal semantics still does not explain the intuition that humans attach to
such programs. For that, we need to study the informal
semantics of the logic. So far,  two informal semantics
have been proposed that can explain  the intuitive meaning of logic
programs.  One is the view of logic programs under stable semantics \cite{iclp/GelfondL88}
as a non-monotonic logic closely related to default logic and
autoepistemic reasoning developed by Gelfond and
Lifschitz~\cite{GelfondL91}. The second is the view of logic programs as
{\bf definitions} of concepts. This view was implicit already in
Clark's work \cite{adbt/Clark78} on completion semantics and in the work by Van
Gelder, Ross and Schlipf \cite{GelderRS91,VanGelder93} on the well-founded semantics. It was
elaborated later in a series of papers
\cite{Denecker98,tocl/DeneckerBM01,KR/DeneckerV14}.


The informal concept of definition (as found in scientific texts) has
several interesting aspects.  First, it is a rule-based linguistic
concept of informal language: definitions, certainly inductive ones,
are commonly phrased as conditionals or sets of these. Second,
definitions are second nature to us. Much human knowledge is of
definitional nature; this includes inductive and recursive
definitions\footnote{In this text, we use the names \emph{recursive definition} and \emph{inductive definition} interchangeably.}.  
Many familiar recursive logic programs are
obviously representations of inductive definitions (e.g., the {\tt
  member}, {\tt append} and transitive closure programs). Third,
definitions are of mathematical precision: they are the building
blocks of formal mathematical theories. Fourth, it is a well-known
consequence of the compactness theorem that definitions, in particular
inductive ones, cannot, in general, be correctly expressed in
classical first-order logic. Fifth, recently
\cite{KR/DeneckerV14}, it was shown that rule sets under two-valued
well-founded semantics correctly formalize all main sorts of informal
definitions that we find in scientific text in the sense that the
interpretation of the defined symbols in the well-founded model of a
rule set always coincides with the set defined by the informal definition
represented by the rule set. All these are solid arguments that the
concept of definition is a good candidate for the informal semantics
of logic programming. Furthermore, they suggest to define a rule-based
logic construct under the well-founded semantics for expressing
definitions. Importantly, given that this form of information cannot
be expressed in classical logic, it makes sense to add such a
construct to classical logic. This idea was carried out for the first
time in \cite{Denecker:CL2000} leading to the logic \foid which forms the
basis of the \idp language. Given that logic programs were originally
seen as a fragment of classical logic, the definition of this logic
was certainly a remarkable turn of events.

A definition is a piece of information. It lays a strict, precise,
deterministic logical relationship between the defined concept and the
concepts in terms of which it is defined. E.g., the definition of
transitive closure of a graph specifies a logical relationship between
the transitive closure relation and the graph. A definition, like all other language constructs in \idp theories, is not a
procedure, it cannot be run, it does not specify a problem. It is a
kind of declarative information. 

The following question now naturally arises: if \idp theories are not
programs, how can they be used to solve computational problems?  The
\idp system, which supports the \idp language, is conceived as a {\em
  \KBS} \cite{iclp/DeneckerV08}. The scientific working hypothesis
underlying the knowledge base paradigm is that many computational
problems can be solved by applying some form of inference to a
specification of information of the problem domain. A \KBS essentially
consists of two parts. On the one hand, a formal declarative knowledge
representation \emph{language} and, on the other hand, a number of
powerful and generic
\emph{inference methods} to solve a broad class of tasks using a knowledge base. 
%

The paradigm is inspired by several observations.
First, \emph{imperative programming} languages allow the direct
encoding of
specialised algorithms, but knowledge about the problem domain is hidden deep within those algorithms. This facilitates high-performance solutions, but makes debugging and maintenance very difficult. 
Second, a program is typically written to \emph{perform one task} and perform it well, but cannot handle many related tasks based on the same knowledge.  
Third, \emph{knowledge representation} languages excel at representing knowledge in a natural, human-understandable format. Programming language designers are starting to realize this and provide constructs to express generic knowledge, such as the LINQ \cite{Pialorsi:2007} data queries in C\# and annotation-driven configuration \cite{Deinum2014}.
Lastly, the above-mentioned progress in automated reasoning
techniques facilitates the shift of the  control burden from
programmer to inference engine ever more.
%
The knowledge base paradigm is an answer to these observations:
application knowledge is modelled in a high-level \KR language and
state-of-the-art inference methods 
are applied to reason on the modelled knowledge.
It has also been demonstrated that, while the \KBS approach cannot yet compete with highly tuned algorithms, the effort to reach an acceptable solution (w.r.t.\ computing time or solution optimality) can be much smaller than that to develop an algorithmic solution~\cite{synthesis/2012Gebser,TPLP/BruynoogheBBDDJLRDV}. 
Furthermore, the declarative approach results in software that is less error-prone and more maintainable.

The \idp system is a state-of-the art \KBS. 
The system is already in existence for several years, but only recently evolved into a \KBS. 
Up until 2012, \idp was a model expansion system (the \logicname{IDP2}
system)\footnote{Given a logical theory and a structure interpreting
  the domain of discourse, model expansion searches for a model of the
  theory that extends the structure.} capable of representing
knowledge in a rich extension of \FO and performing model expansion by
applying its grounder \gidl and its solver \minisatid. Recently, we
have extended it into \emph{the IDP knowledge base framework} for
general knowledge representation and reasoning (referred to as
\logicname{IDP3}); the earlier technology is reused for its model expansion
inference.  The \idp system goes beyond the \KBS paradigm: for a \KBS
to be truly applicable in practical software engineering domains, it
needs to provide an imperative programming interface,
see~\cite{IDPKBS}. Such an interface, in which logical components are
first-class citizens, allows users to handle input and output (e.g., to
a physical database), to modify logical objects in a procedural way
and to combine multiple 
inference methods to solve more involved tasks. In this
chapter, we use \KBS to refer to a three-tier 
architecture
consisting of language, 
inference methods, and procedural integration. The \idp system provides such a procedural integration through the scripting language Lua~\cite{SPE/IerusalimschyFC96}.  The system's name IDP, \emph{Imperative-Declarative Programming}, also refers to this paradigm.

In the work revolving around \idp, we can distinguish between the knowledge representation language and the state-of-the-art inference engines. 
One can \emph{naturally} model diverse application domains in the \idp language; this contrasts with many approaches that \emph{encode} knowledge such that a specific inference task becomes efficient. 
Furthermore, \emph{reuse} of knowledge is central. The \idp language is modular and provides fine-grained management of logic components, e.g., it supports \emph{namespaces}. 
The implementation of the inference engines provided by \idp aims at the reuse of similar functionality (see Section \ref{sec:mx}). This has two important advantages: (i) improvement of one inference engine (e.g., due to progress in one field of research) immediately has a beneficial effect on other engines; (ii) once ``generic'' functionality is available, it becomes easy to add new inference engines.
To lower the bar for modellers, we aim at reducing the importance of clever modelling on the performance of the inference engines. Several techniques, such as grounding with bounds~\mycite{GroundingWithBounds}, function detection~\mycite{functionDetection}, automated translation to the most suitable solving paradigm~\cite{ictai/DeCat13} and automated symmetry breaking~\cite{ictai/DevriendtBMDD12} have been devised to reduce the need for an expert modeller.

The rest of the chapter is structured as follows.  In Section
  \ref{sec:preliminaries}, we present the syntax and semantics of
  \foidaggpft, the logic underlying the system. In Section \ref{sec:idp} we present a high-level overview of the \idp system. 
  In Section \ref{sec:language}, we present the \idp language, a user-friendly syntax for  \foidaggpft language components and procedural control. 
  We present advanced language features and inference methods in Section \ref{sec:advanced}. In Section \ref{sec:mx}, we focus on the inner working of some components of the IDP system. More specifically, we describe the workflow of the optimization inference and how users can control the various parts of the optimization engine.
  Applications, tools and performance are
  discussed in Section \ref{sec:practice}, followed by related work
  and a conclusion.  In the rest of the chapter, we  use
  \idp to refer to the current (2018), knowledge base version of the system.

This chapter is a tribute to David
S.\ Warren. The XSB Prolog system~\cite{jacm/ChenW96} by David
S.\ Warren and his students was the first to support the well-founded
semantics and was a milestone in closing the gap between the
procedural semantics of the SLDNF proof procedure \mycite{SLDNF} and the intuitive
declarative semantics of logic programs as formalized by the
well-founded semantics. In fact, XSB is used internally in the \idp system. In a personal communication, David once told  the authors of this chapter that when he learned about FO(ID) and the \idp system, he was less than thrilled; specifically, he found it ``a crazy idea''.  It is with great satisfaction and gratitude that we have noticed that  he changed his mind as can be seen in his LinkedIn editorial that is devoted to the \idp language \cite{url:linkedinWarren}. It is therefore a great pleasure to contribute this chapter in a book that was initiated to honour his 65th birthday.

\ignore{
Programming has evolved from writing assembler code, that directly addresses the hardware, to using high-level programming languages. The latter is made possible by a large number of automatic compilation steps before reaching that hardware level. With the advent of these higher level languages, more complex tasks, with large numbers of various constraints, are being addressed. Moreover, to accommodate for frequent changes in requirements, the ease of understanding and maintenance of software gain importance. Finally, to automate the many tasks that are currently still solved in a laboriously manual way, it is essential that solving these tasks requires less programming expertise and can more easily be mastered by domain experts. There is evidence of this evolution in many fields, including \textit{verification}~\cite{cacm/MouraB11}, \textit{administration}~\cite{datalog/GreenAK12}, \textit{scheduling}~\cite{lpnmr/Balduccini11}, \textit{data mining}~\cite{iclp/Blockeeletall12}, \textit{robotics}~\cite{kr/Thielscher00} and \textit{configuration}~\cite{ppdp/VlaeminckVD09}.

The foundation of the field of \LP in 1974 with Kowalski's seminal
paper {\em Predicate Logic as a Programming
  Language}~\cite{ifip/Kowalski74} was an important step in that
direction, by giving the Horn-clause subset of predicate logic a
procedural interpretation to use it for programming. Gradually, it
emerged that a logic program is in fact a definition and that the
well-founded semantics~\cite{GelderRS91} is the most natural semantics
to capture the meaning of
definitions~\cite{Denecker98,tocl/DeneckerBM01,KR/DeneckerV14}. The
development of XSB Prolog system ~\cite{jacm/ChenW96} by David S.\
Warren was a milestone in closing the gap between the procedural
semantics of the SLDNF proof procedure and the intuitive declarative
semantics of logic programs as formalized by the well-founded
semantics. The XSB Prolog system~\cite{jacm/ChenW96} was the first to
support the well-founded semantics.  Since that time, tremendous
progress has been made in automated reasoning technology, particularly
in SAT solving and \CP. This has allowed the field of \LP to explore
purer forms of declarative programming, where control is handled by
the solver and the user only has to care about the problem
specification, and for which {\em declarative modelling} is perhaps a
more appropriate term. The best-known exponent of this research is the
field of \ASP~\cite{Baral:2003,cacm/BrewkaET11,synthesis/2012Gebser},
where logic programs are interpreted according to the stable
semantics.

Here we describe the IDP language, a different approach which stays
closer to the origins of logic programming. In layman's terms, what it
offers in addition to XSB Prolog is
(i) 
it is purely declarative (no side effects, no fixed operational
semantics), (ii) more expressivity in the body of rules, supporting
not only conjunction and disjunction, but any first-order logic
formula, 
(iii) the ability to cope with missing information, i.e., with
predicates for which no definition is available and with unspecified
functions, (iv) the use of first-order logic formulas to express
constraints on the domain of discourse in case of missing information,
(v) a full separation between knowledge and the reasoning task, and
the support of several inference methods for performing reasoning
tasks (in XSB Prolog, the only reasoning is deductive querying the logic
program). This does not come without a cost, the ability to cope with
infinite domains is currently quite limited, and it is a major
research challenge to overcome this limitation. 

 IDP integrates inductive definitions (a generalisation of Prolog's
 rules under the well-founded semantics) with \FO formulas to express
 general knowledge about the problem domain. The \idp system that
 supports the \idp language is conceived as a {\em Knowledge Base
   System} (KBS) \cite{iclp/DeneckerV08}.
A \KBS essentially consists of two parts. On the one hand, a
\emph{language} that is   both formal and  natural  
 and, on the other hand, a number of powerful and generic
\emph{inference methods} to solve a broad class of tasks using a knowledge base. 
%

The paradigm is inspired by several observations.
First, \emph{imperative programming} languages allow the direct
encoding of
specialised algorithms, but knowledge about the problem domain is hidden deep within those algorithms. This facilitates high-performance solutions, but makes debugging and maintenance very difficult. 
Second, a program is typically written to \emph{perform one task} and perform it well, but cannot handle many related tasks based on the same knowledge.  
Third, \emph{knowledge representation} languages excel at representing knowledge in a natural, human-understandable format. Programming language designers are starting to realize this and provide constructs to express generic knowledge, such as the LINQ data queries in C\# and annotation-driven configuration.
Lastly, the above-mentioned progress in automated reasoning
techniques facilitates the shift of the  control burden from
programmer to inference engine ever more.
%
The knowledge base paradigm is an answer to these observations:
application knowledge is modelled in a high-level \KR language and
state-of-the-art inference methods 
are applied to reason on the modelled knowledge.
It has also been demonstrated that, while the \KBS approach cannot compete with highly tuned algorithms, the effort to reach an acceptable solution (w.r.t.\ computing time or solution optimality) can be much smaller than that to develop an algorithmic solution~\cite{synthesis/2012Gebser,TPLP/BruynoogheBBDDJLRDV}. 
Furthermore, the declarative approach results in software that is less error-prone and more maintainable.

The \idp system is a state-of-the art \KBS. 
The system is already in existence for several years, but only recently evolved into a \KBS. 
Up until 2012, \idp was a model expansion system (the \logicname{IDP2}
system)\footnote{Given a logical theory and a structure interpreting
  the domain of discourse, model expansion searches for a model of the
  theory that extends the structure.} capable of representing
knowledge in a rich extension of \FO and performing model expansion by
applying its grounder \gidl and its solver \minisatid. Recently, we
have extended it into \emph{the \idp knowledge base framework} for
general knowledge representation and reasoning (referred to as
\logicname{IDP3}); the earlier technology is reused for its model expansion
inference.  The \idp system goes beyond the \KBS paradigm: for a \KBS
to be truly applicable in practical software engineering domains, it
needs to provide an imperative programming interface,
see~\cite{IDPKBS}. Such an interface, in which logical components are
first-class citizens, allows users to handle in- and output (e.g., to
a physical database), to modify logical objects in a procedural way
and to combine multiple 
inference methods to solve more involved tasks. In this
chapter, we use \KBS to refer to a three tier 
architecture
consisting of language, 
inference methods and procedural integration. The \idp system provides such a procedural integration through the scripting language Lua~\cite{SPE/IerusalimschyFC96}.  The system's name IDP, \emph{Imperative-Declarative Programming}, also refers to this paradigm.

In the work revolving around \idp, we can distinguish between the knowledge representation language and the state-of-the-art inference engines. 
One can \emph{naturally} model diverse application domains in the \idp language; this contrasts with many approaches that \emph{encode} knowledge such that a specific inference task becomes efficient. 
Furthermore, \emph{reuse} of knowledge is central. The \idp language is modular and provides fine-grained management of logic components. E.g., it supports \emph{namespaces}: formulas and terms can be declared in one component and used in several other components.
The implementation of the inference engines provided by \idp aim at the reuse of similar functionality (see Section \ref{sec:mx}). This has two important advantages: (i) improvement of one inference engine (e.g., due to progress in one field of research) immediately has a beneficial effect on other engines; (ii) once ``generic'' functionality is available, it becomes easy to add new inference engines.
To lower the bar for modellers, we aim at reducing the importance of clever modelling on the performance of the inference engines. Several techniques, such as grounding with bounds~\mycite{GroundingWithBounds}, function detection~\mycite{functionDetection}, automated translation to the most suitable solving paradigm~\cite{ictai/DeCat13} and automated symmetry breaking~\cite{ictai/DevriendtBMDD12} have been devised to reduce the need for an expert modeller.

\bart{
The rest of the chapter is structured as follows.  In Section
  \ref{sec:preliminaries}, we present the syntax and semantics of
  \foidaggpft, the logic underlying the system. In Section \ref{sec:idp} we present a high-level overview of the \idp system. 
  In Section \ref{sec:language}, we present the \idp language, a user-friendly syntax for specifying \foidaggpft language components and procedural control. 
  We present advanced language features and inference methods in Section \ref{sec:advanced}. Afterwards, in Section \ref{sec:mx}, we focus on the inner working of some components of the IDP system: the workflow of the optimization inference and how users can control the various parts of the optimization engine.
  Applications, tools and performance are
  discussed in Section \ref{sec:practice}, followed by related work
  and a conclusion.  In the rest of the chapter, we  use
  \idp to refer to the current (2015), knowledge base version of the system.
  }

}

%% file: 2-preliminaries.tex
\section{\foidaggpft, the Formal Base Language}\label{sec:preliminaries}
\newcommand{\comp}{\ensuremath{\mathrm{comp}}}
\newcommand{\vocT}{voc(\theory)}
\newcommand\restr[2]{\ensuremath{\left.#1\right|_{#2}}}

In this section, we introduce the logic that is the basis of the \idp
language. This logic, \foidaggpft, is an extension of first-order logic (\FO)
with inductive definitions, aggregates, partial functions and types.




\subsection{First-Order Logic}\label{sec:fo}

A \emph{vocabulary} \voc consists of a set of predicate and function symbols,
each with an associated \emph{arity}, the number of arguments they
take.  We sometimes use $P/n$ ($f/n$) to denote the predicate symbol
$P$ (function symbol $f$) with arity $n$.   

A \emph{term} is a variable or an $n$-ary function symbol applied to
$n$ terms.
An \emph{atom} is an $n$-ary predicate symbol applied to $n$ terms.
An atom is a \emph{formula}; if $\f$ and $\f'$ are formulas and $x$ is a variable, then $\lnot \f$, $\f \land \f'$, $\f \lor \f'$, $\forall x: \f$ and $\exists x: \f$ are also formulas.
The expressions $\f \limplies \f'$, $\f \limpliedby \f'$ and $\f \lequiv \f'$ are (as usual) shorthands for $\lnot \f \lor \f'$, $\f \lor \lnot \f'$ and $(\lnot \f \lor \f')\land (\f \lor \lnot \f')$ respectively.
A \emph{literal} (often denoted $l$) is an atom $a$ or its negation $\lnot a$.
A \emph{sentence} is a formula without free (unquantified) variables.
A \emph{theory} \theory over a vocabulary \voc consists of a set of sentences with symbols in \voc.
\renewcommand{\vocT}{{\color{red}DELETE ME}}
A term $t$ containing occurrences of a term $t'$ is denoted as $t[t']$; the replacement of $t'$ in $t$ by $t''$ is denoted as $t[t' \subs t'']$ (similarly for formulas).

A \emph{two-valued structure} (in the literature, sometimes also referred to as an \emph{interpretation}) \struct over a
vocabulary \voc consists of a \emph{domain}
$D$ 
and an interpretation for all symbols in \voc; we use $s^\struct$ to
refer to the interpretation of a symbol $s$ in \struct.  A two-valued
interpretation $P^\struct$ of a predicate symbol $P/n$ is a subset of
$D^n$; a two-valued interpretation $f^\struct$ for a function symbol
$f/n$ is a mapping $D^n\to D$.
The latter mapping can also be represented by a subset of $D^{n+1}$ in
which there is a functional dependency from the first $n$ arguments to
the last one.
Given a domain $D$, a \emph{domain atom} is a tuple $(P,\ddd)$ where $P$ is an $n$-ary predicate symbol and $\ddd\in D^n$ is an $n$-tuple of domain elements. Sometimes, we abuse notation and write a domain atom as $P(\ddd)$.

%
While the domain of standard \FO is unordered, it is often convenient
to assume that there is a total order of the set of all domain elements and that a vocabulary
includes, by default, the binary comparison predicates $=$/2,
$\neq$/2, $<$/2, $>$/2, $\geq$/2 and $\leq$/2; their interpretation is
fixed in accordance with the total order.
%
%

By evaluating a term or formula in a structure \struct, we obtain its
\emph{value}. The value of a term is a domain element, the value of a
formula is a truth value, either true, denoted \ltrue, or false,
denoted \lfalse, hence an element of the set $\{\ltrue,\lfalse\}$.
The value of a term $t$, denoted as $t^\struct$, is $d$ if $t$ is of
the form $f(\ttt)$ and $ f^\struct(\ttt^\struct) = d$.  The value
$P(\ttt)^\struct$ of an atom $P(\ttt)$ in \struct is
\ltrue if $\ttt^\struct \in P^\struct$ and  \lfalse
otherwise. We define $(\f \land \f')^\struct = \ltrue$ if $\f^\struct
=\f'^\struct =\ltrue$, $\f \lor \f'^\struct = \ltrue$ if either
$\f^\struct =\ltrue$ or $\f'^\struct =\ltrue$, $\lnot
\f^\struct=\ltrue$ if $\f^\struct=\lfalse$; $(\forall x: \f)^\struct
=\ltrue$ if $(\f[x\subs d])^\struct=\ltrue$ for all $d\in D$,
$(\exists x: \f)^\struct=\ltrue$ if $(\f[x\subs d])^\struct=\ltrue$ for
at least one $d\in D$.  
%
In the two quantified forms, the replacement of $x$ by $d$ in \f  means that $x$ is interpreted as $d$ when deriving the value of \f in $I$.
We say a two-valued structure $I$ is a \emph{model} of a formula \f or
$I$ \emph{satisfies} \f, denoted as $I \models \f$, if $\f^I = \ltrue$.
Given two tuples $\ttt=(t_1,\dots,t_n)$ and $\ttt'=(t_1',\dots,t_n')$ of terms of equal length $n$, $\ttt = \ttt'$ denotes the conjunction $t_1=t'_1 \land \dots \land t_n=t'_n$.
For vocabularies \voc and \voc' with $\voc' \supseteq \voc$ and a
structure  $I$ over $\voc'$, $\restr{I}{\voc}$ denotes the restriction
of $I$ to the symbols in $\voc$.

Unless the context specifies it differently, \f denotes a formula, $t$
a term, $D$ a domain, \struct a two-valued structure, \I a partial
structure (introduced below), $d$ a domain element, $x$ and $y$
variables, and $\sim$ any comparison predicate.  

Sometimes, it is convenient to use $\mathit{true}$ and $\mathit{false}$ as atoms
in a formula. Therefore, we include them as nullary predicates in every vocabulary. In every
structure \struct, $\mathit{true}$ is interpreted as $\{()\}$, i.e., as the set containing only the empty tuple, hence
$\mathit{true}^\struct = \ltrue$, and $\mathit{false}$ is interpreted as the empty set $\emptyset$, 
i.e., $\mathit{false}^\struct= \lfalse$.

\paragraph{Partial Structures}
A typical problem solving setting is that of model
expansion~\cite{MitchellT05} where one has partial knowledge about a
structure and where the goal is to expand this partial information into
a structure that is a model of the given theory. Hence we also use
partial structures.
A partial structure  over a vocabulary \voc consists
of a \emph{domain} $D$, and a partial interpretation \I of the symbols
in \voc.
With $s^\I$, we refer to the interpretation of a symbol $s$ in a
partial structure \I. 

Whereas the two-valued interpretation of a predicate $P/n$ was defined
as a subset of $D^n$, it can as well be defined as a mapping from
$D^n$ to the set $\{\ltrue,\lfalse\}$. The latter form is better
suited for generalization. The partial interpretation of a predicate
$P/n$ is defined as a mapping from $D^n$ to the set
$\{\ltrue,\lfalse,\lunkn\}$, with \lunkn standing for ``unknown''.
This mapping partitions $D^n$ in the set of true tuples, denoted
$P^\I_{ct}$ (here $ct$ stands for \emph{certainly true}), the set of false tuples, denoted $P^\I_{cf}$ ($cf$ stands for \emph{certainly false}), and the set
of unknown tuples, denoted $P^\I_{u}$. Note that two of these sets
fully determine the partial interpretation of $P$.


Whereas the two-valued interpretation of a term is a single domain
element, this is no longer the case for a partial interpretation. The
partial interpretation of a function $f/n$ is as a mapping
from $D^{n+1}$ to $\{\ltrue,\lfalse,\lunkn\}$. As for predicates, we
can distinguish true tuples $f_{ct}^\I$, false tuples $f_{cf}^\I$, and
unknown tuples $f_{u}^\I$. While a functional dependency holds in the
set $f_{ct}^\I$, this is not the case for the latter two
sets. However, $(\ddd,d) \in f_{ct}^\I$ iff $(\ddd,d') \in f_{cf}^\I$
for all $d' \in D \setminus \{d\}$.

If $\I$ is a partial structure, $U$ a set of domain atoms, and $v$ a truth value, we use $\I[U:v]$ to refer to the partial structure that equals $\I$ except that for each domain atom $P(\ddd)\in U$, it holds that $P(\ddd)^{\I[U:v]}=P^{\I[U:v]}=v$.


The partial structure \I that corresponds to a two-valued
v \struct is such that, for predicate symbols $P/n$,
$P_{ct}^\I=P^I$, $P_{cf}^\I=D^n\setminus P_{ct}^\I$ and, $P_{u}^\I=
\emptyset$ and, for function symbols $f/n$, $f_{ct}^\I= \{(\ddd,d)
\mid f^\struct(\ddd)=d\}$, $f_{cf}^\I= D^{n+1}\setminus f_{ct}^\I$ and
$f_{u}^\I=\emptyset$.

{The \emph{truth order} $<_t$ on truth values is induced by $\lfalse <_t\lunkn<_t\ltrue$.  The \emph{precision order} $<_p$ on truth values is induced by  $\lunkn
<_p \ltrue$ and $\lunkn <_p \lfalse$.
%
%
This order is extended to a precision order over partial
structure.  A partial structure $\I$ is less precise than a
partial structure $\I'$ (notation $\I\leqp\I'$) if, for all
symbols $s\in \voc$, $s_{ct}^\I \subseteq s_{ct}^{\I'}$ and $s_{cf}^\I
\subseteq s_{cf}^{\I'}$. Maximally precise partial structures are
two-valued.

In the remainder of the chapter, partial interpretation or structure is
intended when interpretation or structure is used.

\subsection{Partial Functions}\label{sec:pf}

In standard logic, function symbols denote total functions.  In
practice, partial functions are unavoidable, e.g., a function that
maps persons to their spouse is naturally undefined for singles as
well as for objects that are not a person, and the arithmetic
operation division is undefined when the denominator is zero.
So, our logic supports partial functions; however, defining a
semantics for partial functions gives rise to undefined terms (also called non-denoting); this is a subject of
controversy \cite{UndefinednessInConstraintLangs,phd/Wittocx10}.

The simplest solution is to restrict the syntax of formulas. One
could, e.g., only allow terms of the form $f(t)$ in contexts where it
is certain that $f(t)$ is defined. This option is often taken in
mathematics, where terms like, e.g., $1/0$ 
are considered meaningless, but quantifications of the form $\forall
x: x\neq 0\limplies 1/x \neq 42$ 
are allowed as it is clear that the division $1/x$ 
will be defined for all relevant $x$.  This idea has been implemented
for example in the Rodin toolset for Event-B
\cite{journals/sttt/AbrialBHHMV10}, where for every occurrence of a
partial function, it should be provable that the function will only by
applied to values in its domain.
However, this approach is too restrictive for a \KBS. For example, in
planning problems, the function $Do/1$ in a term $Do(t)$ that refers to
the action performed at time $t$ is typically a partial one. It can be
pretty hard to come up with the right $\mathit{Condition}/1$ predicate
such that one can write $\forall t: \mathit{Condition}(t) \limplies
\ldots Do(t) \ldots$. It is entirely impossible, when the partial
function is a constant for which it is a priori unclear whether it is
defined, such as $\mathit{Unicorn}$. So we allow terms in contexts
where they can be undefined. This brings us to the question what an
ambiguous form such as $\mathit{White}(\mathit{Unicorn})$ means. Does
it mean ``if the unicorn exists then it is white'', i.e., $\forall x:
\mathit{Unicorn}=x \limplies \mathit{White}(x)$ or ``the unicorn
exists and is white'', i.e., $\exists x: \mathit{Unicorn}=x \land
\mathit{White}(x)$ (which equals $(\exists x: \mathit{Unicorn}=x)
\land \mathit{White}(\mathit{Unicorn})$). For the user, having to
write the longer unambiguous form is rather inconvenient (especially
in case of nested partial functions).



The current approach, which is the result of some years of
experimenting, is based on the \emph{relational semantics} proposed in
\cite{UndefinednessInConstraintLangs}. It turns out to be flexible,
intuitive and to allow for elegant modelling. The ambiguous form
$A(f(\ttt))$ is given the second of the above two meanings, namely $
\exists x: f(\ttt)=x \land A[x]$ or equivalently $ \exists x:
f(\ttt)=x \land A[f(\ttt)]$. When the user is in doubt or prefers the
other form, he should avoid the ambiguous form and explicitly write
one of the explicit forms.

In a two-valued structure \struct, the value of a partial function
$f/n$ is a mapping $f^\struct: S\to D$ where $S$ is some subset of
$D^n$. 
In a partial structure \I, $f/n$ is undefined for \ddd if and only if $(\ddd,d) \in f^\I_{cf}$ for all $d \in D$.
We say that the image $f(\ddd)$ is \emph{undefined} when $f$ is
undefined for $\ddd$.  The interpretation of a term with a direct
subterm that is undefined is also undefined; that of an atom with a
direct subterm that is undefined is \emph{false}. This corresponds to
the above described semantics.

\subsection{Arithmetic}

Standard FO can easily be extended with arithmetic. Indeed, numbers
can be added to the domain and various (partial) functions can be
included in the vocabulary to perform arithmetic.

So far, the \idp system only supports arithmetic over integers. This
is our motivation to include the integers in every domain of
\foidaggpft and the arithmetic partial functions +/2, -/2, -/1, */2
(multiplication), //2 (division) \%/2 (modulo) and $abs/1$ in every
vocabulary\footnote{The current implementation of the \idp
  language, described in Section~\ref{sec:language}, has only limited
  support for integer division.}. To refer to these integers, every
vocabulary also includes the constant symbol $n$ for every integer
$n$.  Furthermore, in every structure, the interpretation of these
integer constants is fixed to the corresponding integer in the domain.
%


%
%
%
%
%
%

%

\subsection{Aggregates}

Aggregates are an important language construct to boost the
expressiveness of first-order logic. \foidaggpft includes the
aggregate functions \emph{cardinality}, \emph{sum}, \emph{product},
\emph{minimum} and \emph{maximum}. The basic underlying concept is the
set expression $\{(\xxx)\mid \f\}$ or $\{(\xxx,t)\mid \f\}$ where $\xxx$ is
a tuple of new variables and $t$ is a term $t[\xxx,\yyy]$ with
variables in $\xxx$ and in the free variables $\yyy$ of the expression. Given
a two-valued structure \struct, such a set expression denotes the
set of tuples $\{(\xxx)^\struct \mid \f^\struct = \ltrue\}$ or
$\{(\xxx,t)^\struct \mid (\exists z: z=t \land \f)^\struct = \ltrue\}$, where the existential quantification ranges over the domain of $t$.
The inclusion of $\exists z : z=t$ in the set prescription
disambiguates the set in case of partial functions in \f. 

Cardinality expressions are written in \foidaggpft as $\#\{\xxx : \f\}$ and this term
denotes the number of tuples in the set. For the other four, the
aggregate expressions take the form $sum\{(\xxx,t) : \f\}$,
$prod\{(\xxx,t) : \f\}$, $min\{(\xxx,t) : \f\}$, and $max\{(\xxx,t) :
\f\}$ respectively. These expressions denote the value of the
aggregate function on the multiset obtained by extracting the last
element $t^\struct$ of each tuple.
For instance $sum\{((x_1,x_2),x_2) : P(x_1,x_2)\}$ sums the values of the second element of all tuples in $P$. 
If there are multiple occurrences of the same ``second element'' in tuples of $P$, then they are counted according to their multiplicity. 

All aggregate functions are partial functions; they are   undefined
for infinite sets.  Moreover, all aggregate functions except  \#  are only defined for sets containing a tuple
$(\dots,d)$ with $ d$ an integer. The aggregates $min$ and $max$
are also undefined for the empty set; in contrast, $sum$ and $prod$
map the empty set to 0 and 1 respectively. 

The aggregates supported by \foidaggpft also include the class of
formulas $\exists_{\sim n}x: \f$ with $n$ a natural number and $\sim$
one of the comparison operators. While such formulas are equivalent
with $\#\{x :\f\}\sim n$, they are more concise and a convenient
extension of existensial quantification.  They allow one to express
``there exists exactly $n$'' ($=n$), ``at most $n$'' ($\leq n$),
``less than $n$'' ($<n$), ``more than $n$'' ($>n$), ``at least $n$''
($\geq n$), and ``there does not exist exactly $n$'' ($\neq n$) values for $x$
such that \f holds. Note that $\exists_{\geq 1}x: \f$ is equivalent
with $\exists x: \f$, and $\exists_{= 0}x: \f$ with $\neg(\exists
x: \f)$.

\subsection{Definitions}\label{subsec:def}

The logic \foidaggpft contains a definition construct to express
different kinds of (possibly inductive) definitions. This construct is
one of the most original aspects of \foidaggpft and we explain it in
more detail. For additional details we refer to
\cite{tocl/DeneckerT08}.

Definitions are important building blocks of any scientific theory.
However, there is no general way to express inductive/recursive
definitions in FO.
%
Though the notion of definition is informal, definitions have some
extraordinary properties. Certainly those used in formal mathematical
text strike us for the precision of their meaning. The formal
semantics of \foidaggpft definitions carefully formalizes this
meaning.  Several types of informal definitions can be distinguished.
Below, the three most common ones are illustrated:
Example~\ref{between} is a non-recursive one, Example~\ref{deftrans}
is a monotone one, while Example~\ref{defsat} is by induction over a
well-founded order, namely over the  subformula
order. 
Definitions over a well-founded order frequently contain
non-monotone rules. For instance the rule defining $I\models
\neg\alpha$ has a non-monotone condition $I\not \models \alpha$.

\begin{example}\label{between}

Let $a,b$ and $c$ be integers; $a$ is between $b$ and $c$ iff $b \leq
a$ and $a \leq c$.

\end{example}

\begin{example}\label{deftrans}  
  Let $(N,E)$ be a graph with nodes $N$ and edges $E$. The transitive
  closure $T$ of $(N,E)$ is defined inductively as follows.
\begin{itemize}
	\item If $(a,b)\in E$, then $(a,b) \in T$,
	\item If for some $c\in N$, it holds that $(a,c)\in T$ and $(c,b)\in T$, then also $(a,b)\in T$. 
\end{itemize}
\end{example}

\begin{example}\label{defsat}
 Let $I$ be a two-valued structure of a propositional vocabulary.
The satisfaction relation $\models$ 
is defined by induction over the structure of formulas:
\begin{itemize}
\item $I\models P$ if $P\in I$.
\item $I\models \alpha\land\beta$ if $I\models \alpha$ and $I\models \beta$. 
\item $I\models \alpha\lor\beta$ if $I\models \alpha$ or $I\models \beta$ (or both).  
\item $I\models \neg\alpha$ if $I\not \models \alpha$. 
\end{itemize}
\end{example}

In \foidaggpft,
a \emph{formal} definition \D is a set of rules of the form $\forall \xxx: P(\ttt) \lrule \f$ or $\forall \xxx: f(\ttt)=t' \lrule \f$, with the free variables of \f and the variables in $\ttt$ and $t'$ amongst the $\xxx$.
We refer to $P(\ttt)$ and $f(\ttt)=t'$ as the \emph{head} of the rule and to \f as the \emph{body}.
In the first form, $P$ is the defined symbol; in the second, $f$ is.
The defined symbols of \D are all symbols that are defined by at least one of its rules; all other symbols occurring in \D are called {\em parameters} or {\em open} symbols of \D. 
Intuitively, for each two-valued structure of the
  parameters, $\D$ determines the interpretation of  the
  defined symbols in a unique way. For instance, the definition of transitive closure can be formalized as follows 
\begin{ltheo}
	\begin{ldef}
		\LRule{\forall a, b: T(a,b)}{E(a,b)}\\
		\LRule{\forall a, b: T(a,b)}{\exists c: T(a,c)\land T(c,b)}
	\end{ldef}

\end{ltheo}

The different sorts of definitions have different semantic
properties. It is commonly assumed that the defined set is the
\emph{least} set that satisfies the rules of the definition, i.e., the
least set such that the head is true whenever the body is
true. However, this is only true for monotone
definitions. 
It does not hold for non-monotone definitions as the following example,
from \cite{KR/DeneckerV14}, illustrates.
\begin{ltheo}
	\begin{ldef}
		\LRule{Even(0)}\\
		\LRule{\forall x: Even(x+1)}{\lnot Even(x)}
	\end{ldef}
\end{ltheo}
Intuitively, this defines the infinite set $\{Even(0), Even(2),
Even(4) \ldots\}$ of even numbers. However, also the infinite set
$\{Even(0),$ $Even(2),$ $Even(3),$ $Even(5),$ $\ldots\}$ satisfies the rules
as for every rule instance with a true body, the head is also
true. Both solutions are minimal; however, none is ``least''.

%
Still, there is an explanation that applies to all kind of definitions
  \cite{BuchholzFPS81}: the set defined by an inductive definition is
  the result of a construction process.
The construction starts with the empty set, and proceeds by iteratively applying non-satisfied rules, till the set is saturated.
In the case of monotone definitions, rules can be applied in any
order; but in the case of definitions over a well-founded order, rule
application must follow the well-founded order. This condition is
necessary for the non-monotone rules. If they would be applied too
early, later rule applications may invalidate their condition. E.g.,
in the initial step of the construction of $\models$, when the
relation is still empty, we could derive $I\models\neg\f$ for each \f,
but the condition $I\not\models\f$ will in many cases later become
invalidated. The role of the induction order is exactly to prevent
such an untimely rule application. 
E.g., to prevent deriving $Even(3)$ before $Even(2)$ has been
derived. 
%


The problem we face in formalizing this idea for the semantics of
\foidaggpft definitions, is that the syntax of \foidaggpft does not
specify an explicit induction order for non-monotone \foidaggpft
definitions. Thus, the question is whether one can somehow ``guess''
the induction order. Indeed, if we look back at the definition of Example~\ref{defsat}, we see that the order is implicit in the structure of the rules: formulas in the head of rules are always larger in the induction order than those in the body. This holds true in general. It should be possible then to design a mathematical procedure that somehow is capable to exploit this implicit structure. 

In \cite{KR/DeneckerV14}, this idea was elaborated. The induction
process of an \foidaggpft definition is formalized as a sequence of
three-valued structures 
of increasing precision. Such a structure 
records what elements have been derived to be in the set, what
elements have been derived to be out of the set, and which have not
been derived yet. Using the current three-valued structure, one
can then establish whether it is safe to apply a rule or not. All
induction sequences can be proven to converge. In case the definition
has the form of a logic program and the underlying structure is a
Herbrand interpretation, the resulting process can be proven to
converge to the well-known well-founded model of the
program~\cite{GelderRS91}. As such, the semantics of \foidaggpft
definitions is a generalization of the well-founded semantics, to
arbitrary bodies, arbitrary structures and with parameters. This
(extended) well-founded semantics provides a uniform formalization for
the two most common forms of induction (monotone and over a
well-founded order) and even for the less common form of iterated
induction~\cite{BuchholzFPS81}. Compared to other logics of iterated
inductive definitions, e.g., the work in ~\cite{BuchholzFPS81}, the
contribution is that the order does not have to be expressed;
a substantial advantage as this can be very tedious.

The satisfaction relation of \FO is thus extended to handle
definitions by means of the well-founded semantics~\cite{VanGelder93}, since it 
formalizes the informal semantics of rule sets as
inductive
definitions~\cite{Denecker98,tocl/DeneckerBM01,KR/DeneckerV14}.  We
now formally describe how this is done.  First, consider definitions
\D that define only predicate symbols. We use the parametrised well-founded semantics. This semantics has been implicitly present in the literature for a long time, by assigning a meaning to an \emph{intensional} database. 
We follow the formalisation
by Denecker and Vennekens \cite{lpnmr/DeneckerV07}.
We say that a two-valued structure \struct satisfies \D ($\struct\models \D$)
if \struct is the parametrised well-founded model of \D, that means
that \struct is the well-founded model of $\D$ when the open
symbols/parameters are interpreted as in \struct.

Checking the latter is done by computing the well-founded model of \D. This can be computed as the limit of a \emph{well-founded induction} \cite{lpnmr/DeneckerV07}, defined below. 
\begin{definition}[Refinement]
We call a partial structure $\I'$ a \emph{refinement} of partial structure \I if one of the following holds:
\begin{itemize}
 \item for every defined predicate $P$ and every tuple of domain elements $\ddd$, \[P(\ddd)^{\I'}=\max_{\leq_t}\{\varphi[\xxx/\ddd]^\I\mid \forall \xxx: P(\xxx)\lrule \varphi  \text{ is a rule in } \D\}.\]
 \item $\I'=\I[U: \lfalse]$, where $U$ is a set of domain atoms unknown in $\I$ such that for every $P(\ddd)\in U$ and every rule $\forall \xxx: P(\xxx)\lrule\varphi$ in \D, it holds that $\varphi[\xxx/\ddd]^{\I'}=\lfalse$ (such a set $U$ is called an unfounded set of $\D$ in \I \cite{GelderRS91}).
\end{itemize}
A refinement is \emph{strict} if $\I'\neq\I$.
\end{definition}

The first refinement evaluates all rule bodies of all defined domain
atoms and assigns the largest truth value (e.g., if one rule derives an atom to be true, and the second rule does not yet derive information about that atom ($\lunkn$), the atom obtains the value $\ltrue$) to each defined atom. The second refinement
identifies an unfounded set: a set of domain atoms such that the bodies of rules defining them can only become true if at least one of these atoms is true in the first place (due to cyclic dependencies). Such atoms can never be derived constructively using the definition, hence they must be false.

\begin{definition}[Well-founded induction]
Let $\I$ be a partial structure that interprets only the open symbols of \D. A \emph{well-founded induction} of $\D$ in \I is a sequence $(\I_i)_{i\leq n}$, with $n\in \nat$, of partial structures such that the following hold:\footnote{In the infinite case, a similar sequence can be constructed. For details, see \cite{lpnmr/DeneckerV07}.}
\begin{itemize}
 \item $\I_0=\I$;
 \item for each $i<n$, $\I_{i+1}$ is a refinement of $\I_i$.
\end{itemize}
A well-founded induction is \emph{terminal} if its limit ($\I_n$) has no strict refinements.
\end{definition}
Denecker and Vennekens  \cite{lpnmr/DeneckerV07} showed  that all terminal well-founded inductions in \I have the same limit, namely the well-founded model of \D in context \I.




To extend this satisfaction check to definitions defining functions,
one treats a function $f/n$ as if it is defining an $n+1$-ary
relation.
For what concerns the use of partial functions in the head of a rule,
note that $\forall \yyy: p(\ttt) \lrule \f$ is equivalent with
$\forall \xxx \yyy: p(\xxx) \lrule \xxx=\ttt \land \f$.  Partial functions
in bodies have the same meaning as in formulas.

In \foidaggpft, a definition is seen in a pure declarative way, as a
proposition stating a special logical relationship between defined
predicates and parameter symbols. In case a theory contains multiple
definitions of the same predicate, the theory states multiple
independent such propositions. For instance, a theory that contains
\begin{ltheo}
  \begin{ldef}
    \LRule{\forall x: Human(x)}{Man(x)\lor Woman(x)}
  \end{ldef}\\
  \begin{ldef}
    \LRule{\forall x: Human(x)}{Child(x)\lor Adult(x)}
  \end{ldef}
\end{ltheo}
states that Human is the union of  men and women and also that human is
the union of children and adults. This implies for example that the
union of men and women, and of children and adults is identical.
Note that this declarative view implies that the definition
\begin{ltheo}
  \begin{ldef}
    \LRule{p}{q}\\
    \LRule{q}{p}
  \end{ldef}
\end{ltheo}
has a different meaning than the pair of definitions
\begin{ltheo}
  \begin{ldef}
    \LRule{p}{q}
  \end{ldef}\\
  \begin{ldef}
    \LRule{q}{p}
  \end{ldef}
\end{ltheo}
Indeed, in the former, $p$ and $q$ are false in the only well-founded
model. In the latter, the structure in which $p$ and $q$ are true
is also a model since we now have two definitions, each with a parameter.

To reason with definitions, the \idp solver makes also use of
  their completion.
The \emph{completion} of \D for a symbol $P$, defined in \D by the rules $\forall \xxx_i: P(\ttt_i) \lrule \f_i$ with $i \in [1,n]$, is the set consisting of the sentence $\forall \xxx_i: \f_i \limplies P(\ttt_i)$ for each $i\in [1,n]$ and the sentence $\forall \xxx: P(\xxx) \limplies \bigor_{i \in [1,n]} (\xxx=\ttt_i \land \f_i)$; the completion for defined function symbols is defined similarly.
This set is denoted as $comp_{P,\D}$, the union of all these sets for \D as $comp_\D$. 
It is well known (see, e.g., \cite{GelderRS91}) that if $I\models \D$ then $I\models \comp_{\D}$ but not always vice-versa (e.g., the inductive definition expressing transitive closure is stronger than its completion).

\subsection{Types} \label{sec:preliminaries:types}

While first-order logic is untyped, the real world is typed. For
example, in a course scheduling problem, we can distinguish persons,
which can be further divided in teachers and students, courses, rooms,
time slots, etc. The advantages are well known in the field of programming
languages. For example, the introduction of the book of Pierce
\cite{MIT/Pierce2002} lists, among others, early detection of errors
and a minimum of documentation. 
These advantages also hold for a knowledge representation language.
Moreover, the use of types leads to more detailed and more accurate
modelling of the different types of objects in the domain of discourse.


\foidaggpft is a simple \emph{order sorted logic}. This means that a vocabulary contains a
set of types on which a subtype relation is defined. The corresponding
type hierarchy is a set of trees. Two types are \emph{disjoint} if they have
no common supertype (a type is a supertype of itself). Every
vocabulary includes the type $\mathit{int}$ of the integers and its
subtype $\mathit{nat}$ of the natural numbers.  All predicate and
function symbols of a vocabulary are typed by means of a type
signature which associates a type with each argument position, and, in
the case of functions, with the result. All variable occurrences in a
theory are typed; all occurrences of a variable within the same scope
have the same type; the type of a variable is given when it is
introduced in a formula or set expression, for example, $\forall
x[T]:\f[x]$ and $\#\{ x[T]~y[T']:\f[x,y]\}<3$.



In a structure, a domain $D_T$ is associated with each type $T$;
for a subtype $T_1$  of $T_2$, $D_{T_1}$ is a subset of $D_{T_2}$; if $T_1$ and $T_2$ are disjoint types, $D_{T_1}$ and $D_{T_2}$ must be disjoint. Structures are well-typed. This means that for a predicate symbol $P$ of type $(T_1,\dots,T_n)$, its value $P^\struct$ belongs to the Cartesian product $D_{T_1}\times\dots\times D_{T_n}$, and for a function symbol $f$ from $(T_1,\dots,T_n)$ to $T$, its value $f^\struct$ is a partial function from $D_{T_1}\times\dots\times D_{T_n}$ to $D_{T}$. For the evaluation of quantified
formulas $\forall x[T]: \f$ and $\exists x[T]:\f$, enumeration of the values of $x$ is 
over the domain $D_T$. Similarly, in a set expression
$\{(\xxx\,[\TTT],t)| \f\}$ of an aggregate, each $x_i$ is assigned
domain elements from $D_{T_i}$.
Note that a term $t$ of type $T_1$ that occurs in an argument of an
atom or function where a term of type $T_2$ is expected can have the
meaning of an undefined term (Section~\ref{sec:pf}). Indeed, the
atom evaluates to false or the function is undefined when $t$
evaluates to a domain element outside $D_{T_2}$. While this cannot
happen when $T_2$ is a supertype of $T_1$, it always is the case when
$T_1$ and $T_2$ are disjoint; it depends on the evaluation when
there is a third type that is a supertype of both. When the types are
disjoint, it is appropriate to raise a type error as this is likely a
design error in the logical theory.

%% file: 3a-idp.tex
\section[<toc-entry>]{\idp as a Knowledge Base System}\label{sec:idp}

We start the section with a description of the architecture and a
discussion of design decisions. We finish with sketching an
application where the same knowledge is used for different tasks.

\subsection{Architecture and Design Decisions}

Here, we introduce and motivate the basic design decisions underlying the
  \idp system, the decisions that determine the look and feel of \idp
  as a \KBS. \idp is an implementation of a \KBS.  Besides the two main parts,
  the \emph{declarative language} and the \emph{inferences methods},
  there is also a part that provides \emph{procedural integration}.
  An overview is shown in Figure \ref{fig:idpkbs}.

\input{image_kbs}

The first design decision, the one most visible to users, is
  about the language of the \KBS. The language should be (i)
  \emph{rich} enough so that users can express all their needs; (ii)
  \emph{natural} enough so that theories stay close to the original
  (natural language) problem statement and are easy to read and to
  debug; and (iii) \emph{modular} enough to allow for reuse and future
  extensions.

It is sometimes argued that the expressiveness of a language should
  be limited, to avoid that the language becomes undecidable or
  intractable.
  We disagree. First, note that decidability and
  tractability depend on the task at hand. While deduction in first
  order logic is undecidable, other forms of inference, such as model
  expansion and querying in the context of a finite domain, are
  decidable. Second, while a more expressive language might allow users to
  express tasks high in the polynomial hierarchy, that does not imply
  that simple tasks become harder to solve. Rather to the contrary,
  stating the problem in a richer language sometimes allow the \KBS to
  exploit structural information that would be hidden in a more lower-level
  problem statement.

To address the requirement of a rich and a natural language, we have opted for \foidaggpft, FO extended with definitions, aggregates, partial functions and types. 
We choose for first-order logic because conjunction, disjunction, universal and existential quantification have very natural meanings. 
Extensions are needed because FO has various weaknesses. Inductive definitions overcome the weakness that FO cannot express inductively defined concepts. Also non-inductive definitions are very useful. 
In mathematical texts, it is common practice to use ``if'' when defining concepts; this ``if'' here is (in natural language) a  
 \emph{definitional implication}. 
Aggregates allow users to concisely express information requiring lengthy and complex FO formulas. 
Types are omnipresent in the context of natural language, where quantification typically refers to a specific set of objects (e.g., everyone is mortal).
For integration with a procedural language, \idp 
 currently offers an interface to the languages Lua and C++.

The third language requirement, modularity, is important both at the language and at the system level. An advantage of first-order logic as basis of our language is  that language extensions can be added without much
interference at the syntactical and semantical level.
   For example to  introduce aggregates to FO, it suffices  to extend the satisfaction  of atoms in which an aggregate occurs in order to obtain a semantics for a language extended with aggregates.

On the system level, 
  we have also attempted to organize inference engines in a modular way so that components can be reused in multiple engines. 
  For example, the model expansion inference is currently implemented as ground-and-solve; the solver can be used separately from the grounder, and the grounding phase is composed of several smaller, reusable parts (for example, evaluation of input-*-definitions~\cite{tplp/Jansen13}). 
  Also various approaches to preprocess simple theories in order to improve their computational efficiently are integrated in the system. 
  Examples are symmetry detection \cite{sat/DevriendtBBD16}  and symmetry breaking \cite{ictai/DevriendtBMDD12}
  methods  and the use of deduction to
  detect functional dependencies \mycite{functionDetection}.
 Such preprocessing techniques also improve the user-friendliness and robustness of the \KBS as a whole. Indeed, they let a user focus on the declarative modelling,   and  partly relieve the user from the task of fine-tuning it on a specific solver.  

\subsection{Multiple Inference Methods Within One Application Domain}

Given any knowledge base, there are often multiple applications that require different kinds of inference.
By way of example, we explore the setting of a university course management system.
Its input is a  database with information on students, professors, classrooms, etc.
One task of the system is to help students choose their courses satisfying certain restrictions. 
Such an application is usually interactive; students make
choices and, in between, the system checks the knowledge base. It
removes choices when they become invalid,  adds required prerequisites
when a course is selected,  etc; this is an example of \emph{propagation} inference. When the student has made all choices he deems important, the system could use \emph{the same} knowledge base to complete the student's choices to obtain a complete schedule. This type of inference is called {\em model
  generation} or {\em model expansion}: one starts with partial
information (certain selections have been made) and wants to extend it into a complete solution, namely a model of the course selection theory. 
Another task where model expansion is needed in the same application area is to generate a schedule where every course is assigned
a location and a starting time such that 1) no person has to be at two
places at the same time, 2) no room is double-booked and 3)
availability of professors is taken into account.
However, due to the large number of optional courses, such a solution (in which no student has overlapping courses) will probably not exist.
In this case, we might want to find a solution in which the number of
conflicts is minimal; this requires {\em minimization} inference. 
Now, one might want to mail students with schedules with overlaps 
to give them the opportunity to change their selection.
Hence, the solution of the minimization inference should be {\em queried} to find the overlapping courses for every student.
In the course of a semester, professors might have to cancel a lecture
due to other urgent obligations. 
In that case, we want to find a {\em revision} of the current
schedule, taking the changed restrictions into account and 
minimizing the number of changes with respect to the current schedule.
In case such revisions are done manually, the {\em model checking}
inference can be used to ensure that no new conflicts are introduced. If some
conflict does occur, an \emph{explanation} should be provided.
Finally, if a valid schedule is found, a {\em visualization} inference
can be used to create an easy-to-understand, visual representation
of the schedule, personalized by the viewer's status (student,
professor, administrative personnel,  etc.).
Part of such an application, using the \idp system as a back end, is shown at \url{http://krr.bitbucket.org/courses}.

%
%

%% file: image_kbs.tex
\tikzstyle{block} = [draw=red, fill=blue!20, rectangle, rounded corners, inner sep=10pt, inner ysep=20pt]
\tikzstyle{mainblock} = [draw=blue, fill=blue!20, very thick, rectangle, rounded corners, inner sep=10pt, inner sep=8pt]
\tikzstyle{fancytitle} =[fill=NavyBlue, text=white]

\pgfdeclarelayer{background}
\pgfdeclarelayer{foreground}
\pgfsetlayers{background,main,foreground}
    
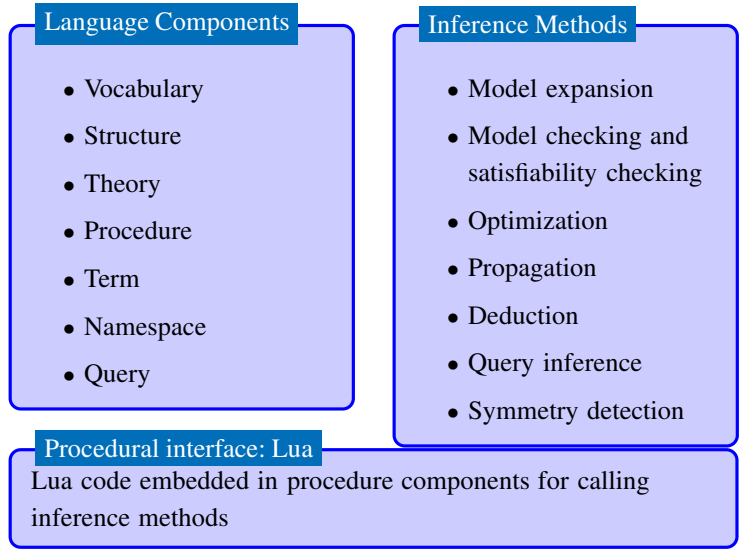
\begin{figure}
\centering
\begin{tikzpicture}[auto, every text node part/.style={align=left}, node distance = 0.5cm]
	\node [mainblock, text width=4cm] (knowledge) { 
		\begin{itemize}
		  \item Vocabulary
		  \item Structure
		  \item Theory
		  \item Procedure
		  \item Term
		  \item Namespace
                  \item Query
		\end{itemize}
	};

	\node [mainblock, right=of knowledge.north east, text width=4cm, anchor = north west] (inferences) {
		\begin{itemize}
		  \item Model expansion
		  \item Model checking and satisfiability checking
		  \item Optimization
                  \item Propagation
		  \item Deduction
		  \item Query inference
		  \item Symmetry detection
		\end{itemize}
	};
	
	\node [mainblock, below=of knowledge.south west, text
        width=9.1cm, anchor = north west] (proceduralint) {
                Lua code embedded in procedure components for calling inference methods
	};
	
	\node[fancytitle, right=10pt] at (proceduralint.north west) { Procedural interface: Lua};
	\node[fancytitle, right=10pt] at (inferences.north west)
        {Inference Methods};
	\node[fancytitle, right=10pt] at (knowledge.north west)
        {Language Components};
\end{tikzpicture}
\caption{High-level representation of a KBS system}
\label{fig:idpkbs}
\end{figure}

%% file: 3b-language.tex

\section[<toc-entry>]
{The \idp Language}
\label{sec:language}

 The \idp language is the input language of the \idp
  system. A program in the \idp language consists of declarative
  and imperative components. The declarative components are \emph{vocabulary}, \emph{structure},
  \emph{theory} and \emph{term} components. Together, they provide a concrete computer-readable syntax for \foidaggpft. The imperative components
  allow one to perform computational tasks. They consists of
  \emph{procedures}. Each procedure embeds a piece of imperative
  Lua~\cite{SPE/IerusalimschyFC96} code; besides performing standard
  imperative operations, procedures can apply inference methods upon
  \foidaggpft theories encoded in the declarative components. 


Vocabulary, structure and theory components are described in
Section~\ref{sec:logic}; procedure components in
Section~\ref{sec:procedure} and term components in
Section~\ref{sec:term}. In this section we do not strive for
completeness but focus on what is needed to get started using the
\idp system and on providing answers to the difficulties a starting
user might have.


Before describing the different kinds of components, we first discuss a
few general notational conventions. Names are everywhere, they are
used for types, predicates, functions (including constants), and
variables as well as for domain elements in structures. To distinguish
them from numbers, they start with a Latin letter (upper or lower
case) and consist of a sequence of Latin letters and digits; also a
few special characters such as ``\il{\_}'' are allowed. For domain
elements, one can deviate from this convention by using a string
notation.  For details, we refer to the manual \cite{url:idp3manual}.  Comments that fit
on a single line start with ``\il{//}''. One can start longer
comments with ``\il{/*}'' and end them with ``\il{*/}''.

\subsection{The Logic}\label{sec:logic}

\subsubsection{Vocabulary}
The vocabulary of an \foidaggpft theory is represented as a vocabulary
component. We start with an example.

\begin{lstlisting}
vocabulary courses {
  type course	
  type person
  type student isa person
  type instructor isa person
  type age isa nat
  takes(student, course)
  hasAge(person):age
  Vacation
  Boss: person
}
\end{lstlisting}

A vocabulary declaration takes the form ``\il{vocabulary} $\langle$\il{vocname}$\rangle$ \{ $\langle$ typed symbol list$\rangle$\}''. It specifies
the name of the vocabulary, here \il{courses} and its symbols. The
symbol list comprises type symbols, and typed predicate and function symbols.  Each symbol is to be declared on a new
line. Types are declared using the keyword ``\il{type}''. 
A type may be declared as a subtype using  the keyword
``\il{isa}'' followed by a comma-separated list of supertypes.  For example 
``\il{type} A \il{isa} B, C'' declares $A$ as a direct subtype of $B$ and $C$. The declared \il{isa} graph needs to be  acyclic.
 The integers, type \il{int}, and
its subtype, the natural numbers, type \il{nat}, are part of every
vocabulary and need not be declared. The same holds for predicates and
functions that are part of every \foidaggpft theory such as comparison
predicates and arithmetic functions.  Predicates and functions are
introduced by declaring their signature. In the example, \il{takes} is a relation over \il{student}
and \il{course} and \il{hasAge} is a function from \il{person} to
the subtype \il{age},  a subtype of nat. The symbol \il{Vacation} is a propositional symbol, and \il{Boss}  a constant symbol. Partial
functions are introduced by the keyword \il{partial}, for example
\il{partial hasAge(person):age} would declare \il{hasAge} as a partial
function. The \idp system cannot yet cope well with infinite types, so
\il{int} and \il{nat} can better be avoided in signatures of
predicates and functions.

\subsubsection{Structure}
A structure component describes a partial structure for a vocabulary,
in particular the domains of the user declared types. We start again
with an example.

\begin{lstlisting}
structure data1: courses{
  course = {Logic; Math}
  student = {John; Bob; Alice}
  instructor = {Marc; Gerda; Maurice}
  person = {John; Bob; Alice; Marc; Gerda; Maurice}
  age = {1 .. 65}
  takes<ct> = {John,Logic}
  takes<cf> = {Bob,Math}
  hasAge<ct>= {John->25; Bob -> 30; Alice->19}
  Vacation=true
  Boss=Alice
}

\end{lstlisting}
A structure has a name, here \il{data1}, and specifies the  vocabulary that it (partially) interprets, here \il{courses}. It specifies an assignment of values for symbols using  a list of ``symbol=value'' equations. Boolean values are denoted \il{true} or  \il{false}, as illustrated by the equation \il{Vacation=true} in the example. Other base values are numbers, strings, or user-defined domain values like \il{Bob, Math}.  Set values are denoted as ``\{ $\langle$semicolon separated list$\rangle$ \}''. The list may consist of individual entities or of  tuples of individual entities. Tuples are denoted as comma separated lists of domain values, potentially between parentheses "( \dots )". The shorthand ``n..m''  enumerates the integer interval $[n,m]$,  as illustrated  for \il{age}. The same holds for characters, e.g., \il{student = \{a..d\}} is a shorthand for \il{student = \{a;b;c;d\}}.


A structure specifies, implicitly or explicitly, the value of each
type of the vocabulary\footnote{Autocompletion may derive missing type
  domains; e.g., in the absence of a domain for age, autocompletion
  will derive \il{\{19, 25, 30\}} for it, in absence of a domain for
  person, it will derive the union of the \il{student} and
  \il{instructor} domains.}. A value of a type is a set of domain
elements. User-defined domain elements are identified by symbolic identifiers
(e.g., Bob, Math) but these identifiers are not symbols of the
vocabulary and cannot be used in the
theory.  
One can use integers as names for the domain elements of types, even
if this type is not a subtype of \il{int}. For example \il{student =
  \{1..50\}} introduces 50 students.  Note, these domain elements are
not integers, the domain element 49 of type student is different from
domain element 49 of type integer.

A (total) value for a predicate symbol is a set of domain elements or tuples of it. Alternatively, a structure may  specify a partial value for a predicate symbol \il{P}, as an assignment of a list of certainly true, certainly false and unknown tuples to respectively \il{P<ct>}, \il{P<cf>}, and \il{P<u>}. Only two out of   three need to be specified. This is illustrated by \il{takes}.

The value or partial value for functions is specified in an analogous way, with the difference that the user is allowed (but not obliged) to specify a tuple ``{(a1,\dots,an,b)}'' of a function in the form ``\il{a1,.., an -> b}''. For partial interpretations of functions, we refer to the discussion on partial
structures in Section~\ref{sec:fo}. For example, \il{hasAge<cf> = \{Marc -> 1; Marc -> 2\}} can be used to express that the age of Marc is neither 1 nor 2.

In \idp one cannot currently use a constant as a bound in a domain
enumeration, as in  \il{student = \{1..nbstudents\}}, where \il{nbstudents} is a constant symbol whose value is specified elsewhere. Another limitation that was already mentioned earlier, is that domain values such as \il{Bob} and \il{Alice}  (identifiers introduced at the right side of ``symbol=value'' equations), are not part of the vocabulary and cannot appear in theories. 

Recall that there exist also many interpreted symbols (e.g., numerical operators and aggregates) whose values are fixed and are implicit parts of all structures. 

\subsubsection{Theory} A theory  component over some vocabulary is  declared  as ``\il{theory} $\langle$theory name$\rangle$ : $\langle$vocname$\rangle$ \{ \dots \}''. For the syntax of formulas and definitions, that of
the formal base language is followed as closely as possible. Formulas
and rules are terminated with a ``\il{.}'' and rules are grouped in
definitions which are put between ``\il{define \{}'' and ``\il{\}}''.
%
%
The following table provides a translation in a more keyboard
friendly notation\footnote{The IDE at \url{http://dtai.cs.kuleuven.be/krr/idp-ide/} visualizes the symbols in the
  syntax of \foidaggpft.}.


\[
\begin{tabular}{c|c||c|c}
\foidaggpft &\idp language & \foidaggpft &  \idp language  \\
\hline
$\land$ & \il{&} & $\geq$ & \il{>=} \\
$\lor$ & \il{|} &  $=$ &  \il{=}  \\
$\limplies$ & \il{=>}  & $\neq$ & \il{\~=} \\
$\limpliedby$ & \il{<=} & $\lrule$ & \il{<-} \\
$\lequiv$  & \il{<=>} &  $\#\{\xxx :\f\}$ & \il{#\{x1 ... xn :} $\f$\il{\}} \\
$\lnot$ &\il{\~} &$sum\{(\xxx,t) : \f\}$&\il{sum\{x1 ... xn :} $\f$\il{: t\}}\\
$\forall$ &\il{!} & $prod\{(\xxx,t) : \f\}$ & \il{prod\{x1 ... xn :} $\f$\il{: t\}}\\
$\exists$ & \il{?} & $max\{(\xxx,t) : \f\}$ & \il{max\{x1 ... xn :} $\f$\il{: t\}} \\
$\leq$ & \il{=<} & $min\{(\xxx,t) : \f\}$ &  \il{min\{x1 ... xn :}
$\f$\il{: t\}}
\end{tabular}
\]

%
By way of example, we show a vocabulary, a structure and a theory for
a small graph problem that formalizes a connected graph over a set of
nodes.

\begin{lstlisting}
vocabulary V{
  type Node
  Forbidden(Node,Node)
  Edge(Node,Node)
  Reachable(Node)
  Root:Node
}

structure S:V{
  Node = {A..D}
  Forbidden = {A,A; A,B; A,C; B,A; B,B; B,C; C,C; C,D; D,D}
  Root = A
}

theory T: V{
  // inductive definition of Reachable
  define {
    Reachable(Root).
    !x [Node]: Reachable(x) <- 
               ?y [Node]: Reachable(y) & Edge(y,x).
  }

  // The graph is fully connected
  !x [Node] : Reachable(x).

  // No forbidden edges
  !x[Node] y[Node] : Edge(x,y) => ~Forbidden(x,y).
}

\end{lstlisting}
The theory contains a definition\footnote{The \il{define} keyword is optional; it emphasizes that the brackets $\{$ and $\}$ are delimiters of a definition.}   of the \il{Reachable} predicate and
two formulas that constrain the solution. All variables are typed with
type \il{Node}; however, these types can be omitted since
\emph{type inference} will derive them from the signatures of the symbols  in which the variables occur.\footnote{If within the same scope a variable appears in argument positions with different types, the inferred type is their least supertype if it exists, otherwise a type error is raised, as explained in
Section~\ref{sec:preliminaries:types}. If no type  can be inferred, e.g., as in \il{!x: x=x}, also then an error is raised.} Observe that this theory  defines \il{Reachable} as the transitive closure of  parameter \il{Edge}. This definition is syntactically similar to a Prolog program, but unlike  a Prolog program, it is here the defined predicate that is known (it is the set of all nodes) and the parameter predicate that is unknown. It illustrates the declarative understanding of  a definition that  expresses a particular logical relationship between the parameters and the defined symbols, and not a way to compute the defined symbols in terms of the parameters. 

\ignore{\maurice{The motivation to distinguish between \foidaggpft and the
  logic components of the \idp language is that the latter
  incorporates a number of design decisions. One such decision is to
  maintain a clear separation between structure and theory. This is
  sometimes inconvenient, and could be an argument to merge theory and
  structure in a single component; then one would need a mechanism to
  distinguish between constants and domain elements. Such alternative
  design decision would result in a different \idp language, while
  preserving the underlying formal logic. } \marc{Het is waar dat we die separation hebben, en dat dit wellicht geen goede designbeslissing was. Maar of dat nu de reden is voor het verschil tussen de theoretische en de concrete taal, dat berijp ik niet. Ik zou liever niet zoveel aandacht willen besteden aan onze slechte architecturale keuzes. Daarom is mijn voorstel dit te schrappen.
}}

\subsection{Procedure}\label{sec:procedure}

A procedure component is a chunk of Lua code \cite{SPE/IerusalimschyFC96} 
encapsulated in the form of an \idp component (a keyword
\il{procedure}, a name, a list of parameters and the chunk of code
between ``\il{\{}'' and ``\il{\}}''). When the IDP system is run, it
calls the procedure \il{main()}. Typically, one will use the \idp system
to do some reasoning on an \foidaggpft theory. Here is a simple example:

\begin{lstlisting}
procedure main(){
	stdoptions.nbmodels = 0
	printmodels(modelexpand(T,S)) 
}   
\end{lstlisting}

The first line, \il{stdoptions.nbmodels = 0}, configures the \idp
system to compute all models (with a positive number $n$, inference is
stopped after $n$ models have been found; default value is 1). The
second line calls \idp's modelexpand procedure with as input arguments the theory $T$ and the structure $S$ and prints these models
(\il{printmodels}). The \il{modelexpand} procedure calls upon the
solver of the \idp system to search for models of $T$ that expand the
input structure $S$ and returns an array of models. To print a single
model, one can select a model and print it,
e.g. \il{print(modelexpand(T,S)[3])} to print the third model. With
the theory and structure as given in the small graph theory of the
previous section, the above procedure prints 24 models, the first one
being:
\begin{lstlisting}
structure  : V {
  Node = { "A"; "B"; "C"; "D" }
  Edge = { "A","D"; "B","D"; "C","A"; "C","B"; "D","A"; "D","B"; "D","C" }
  Forbidden = { "A","A"; "A","B"; "A","C"; "B","A"; "B","B"; "B","C"; "C","C"; "C","D"; "D","D" }
  Reachable = { "A"; "B"; "C"; "D" }
  Root = "A"
}
\end{lstlisting}
Note that the model is written in the syntactic format of  a structure component. The procedures 
\il{modelexpand} and \il{printmodels} are only  two out of many
predefined procedures in \idp. Many procedures provide other forms of inference that can perform
computational tasks  using the knowledge represented by an
\foidaggpft theory. Other procedures serve  to manipulate and create new logical
components, such as vocabularies, structures and theories. We refer to
the online \idp Web-IDE
(\url{http://dtai.cs.kuleuven.be/krr/idp-ide/}) and the \idp
manual
(\url{https://dtai.cs.kuleuven.be/krr/files/bib/manuals/idp3-manual.pdf})
for examples and details.  The main design philosophy is that all
components in an \idp program are first class citizens. They can be used
to perform various reasoning tasks but can also be manipulated by Lua code to
construct different ones. So, it is possible to set up a complete
workflow. 

This methodology in which fine-grained declarative computation steps are mixed in procedures is an exciting novel way of integrating declarative and procedural knowledge. 
This is illustrated by Bruynooghe et al.~\cite{TPLP/BruynoogheBBDDJLRDV}. This is an application in stemmatology, the study of the family
relationships between different manuscripts (hand made copies) of a
text. It sketches a set of procedures that describe the workflow to
analyse a number of texts. For each text, a data set is read and
analyzed and transformed into structures and vocabularies. These are
then combined with the problem vocabulary and theory and, for each so
called feature in the input data of a text, it is checked whether a
model exists. Finally, for each data set, a summary report about all
its features is reported.

\subsection{Term}\label{sec:term}
One of the available forms of inference in \idp is to solve minimization problems. This inference method  takes as input a theory, a partial structure, and a cost term, and outputs one or more models of the theory expanding the partial structure such that the value of the cost term in these structures is minimal (among the set of all models more precise than the partial structure).  The cost term is to be specified as a separate term
component over the same vocabulary as the theory. To
illustrate, we return to our graph problem and introduce a term to
count the number of edges in the solution. 
\begin{lstlisting}
term t: V{
  #{x y: Edge(x,y)}
}

procedure main(){
  stdoptions.nbmodels = 0
  models, optimal, cost = minimize(T,S,t)
  printmodels(models)
  print(optimal)
  print(cost)
}   
\end{lstlisting}
The \il{main} procedure now calls the minimization inference with the
theory \lstinline{T}, the structure \lstinline{S} and the optimization term \lstinline{t} as
input\ignore{\footnote{Currently, only aggregates are supported; however, the
  manual mentions another form for aggregates, one can write
  \il{sum(f(...))} for any term \il{f}. MARC: IK BEGREEP DEZE FOOTNOTE NIET EN HET LEEK ME OOK NIET RELEVANT. }}. With \lstinline{T} and \lstinline{S} as above,
the procedure returns three models, however, it also returns two other
values: whether optimality has been proven and the value of the term
in the optimal solution. The assignment \il{models, optimal, cost=...}
assigns them to different variables. The first value is an array of
models, which can be printed with \il{printmodels}, the other two are
simple values; they can be printed with the standard \il{print}
command. In this example, optimality is reached with a cost of 3.

\section{Advanced Features}\label{sec:advanced}

\subsection{Constructed Types}
In Prolog and ASP, functions and constants have a fixed
interpretation, their Herbrand interpretation. Equivalently, we can
think of them as logics with built-in \emph{unique names} and
\emph{domain closure} axioms (UNA and DCA), i.e., axioms stating that all those values are different and that the domain consists of nothing more than those values, respectively. In \foidaggpft, the axioms
are not present and interpretation of functions and constants is open
as in standard \FO. Both approaches  have their merits, making it useful to integrate the advantages of both. In ASP, there is work to
incorporate open functions \cite{kr/BartholomewL12,kr/Lifschitz12}. In
\foidaggpft, one way to impose UNA and DCA  is to explicitly specify a Herbrand interpretation. This method is illustrated below for the type of  days of the week, using the following declarations in the vocabulary:
\begin{lstlisting}
  Type Day
  monday: Day
  tuesday: Day
  ...
\end{lstlisting}
and in the structure
\begin{lstlisting}
  Day = {''monday''; ''tuesday'' ; ...}
  monday = ''monday''
  tuesday = ''tuesday''
  ...
\end{lstlisting}
There is a way to avoid this cumbersome approach.
The following constructed type declaration in the vocabulary expresses the same information but in a compact way:
\begin{lstlisting}
type Day constructed from {monday, tuesday, wednesday, thursday, friday}
\end{lstlisting}
This statement declares several things at once: it declares the type
\il{Day} and seven constants of this type, it specifies the values for
this type and for all its constants in every structure of the
vocabulary.  Each constant is interpreted by itself.  Here,
constants and domain elements coincide.

Also non-constant constructor  symbols are  supported this way. For example, having types \il{row} and \il{column}, one can
introduce the constructed type of positions on a chessboard by declaring ``\il{type position constructed from {pos(row,col)}}'', and use it in the definition of a unary predicate \il{queen(position)} representing the positions where  a queen stands. 
In theory, this approach also
works for recursive constructors and types such as list of integers:
\il{type list constructed from \{nil ; cons[int, list]\}}. However,
this creates an infinite type and the current \idp solver cannot cope
with such types.







\subsection{Structuring Components}

All components, vocabularies, theories, structures, terms, and
procedures, as we have shown so far, are part of the implicit global
namespace \il{idpglobal}. This namespace also contains all Lua
procedures that are available to the user of the system. When working
on large projects, different people may work on different parts, each
introducing its own components. To integrate such different parts, the
\idp system provides \emph{namespaces}.  A namespace with name
MySpace is declared by
\begin{lstlisting}
 namespace MySpace {
//   content   of   the   namespace
}
\end{lstlisting}
A namespace can contain other namespaces, and any sort of \idp component including vocabularies, theories,
structures, terms, and procedures. Each component has a full name that is determined by the
hierarchy of namespaces it belongs to. This allows users to
disambiguate components with the same name but belonging to different
namespaces. We refer to the manual for details.

Another useful structuring method is to compose a vocabulary from existing ones. For instance, in the following example \il{W} is composed of the symbols of \il{V} and one function \il{coloring/1:1} from vocabulary \il{U}:\footnote{Here, the notation \il{coloring/1:1} means that \il{coloring} has arity one (\il{/1}) and is a function, i.e., has one output argument (\il{:1}).} 
\begin{lstlisting}
vocabulary W {
extern  vocabulary V
extern  U::coloring/1:1
}
\end{lstlisting}
Such an extension construct is not available for structures and
theories but it could be simulated for them by making use of certain
Lua procedures. For this we refer to the list of Lua procedures
described in the manual. 

Worth mentioning is that there is a
\il{factlist} component to initialize a two valued structure with
Prolog or ASP facts. Also, it is possible to call upon Lua procedures
to initialize a structure. 

\subsection{An Output Vocabulary} 

In many problems we are interested only in the values of some subset of symbols. In case multiple solutions are searched, we are interested only in  models having different interpretations of the output symbols. This is achieved by declaring an output vocabulary, say \il{Vout}, and adding it as an extra parameter to the  modelexpand call: 
\begin{lstlisting}
procedure main(){
    print(modelexpand(T,S,Vout)[1])
}
\end{lstlisting}

\subsection{Inference Methods}

So far we have mentioned \emph{model expansion} inference, invoked as
\il{modelexpand(T,S)} (or \il{modelexpand(T,S,Vout)} if there is an output vocabulary) and \emph{optimization} inference, invoked as
\il{minimize(T,S,t)} (or \il{minimize(T,S,t,Vout)}). The system supports several other inference
methods. We discuss the most important ones. For a complete list,
we refer to the manual. 

\textbf{Query inference} takes as input a \emph{query} component that declares a
set expression of the form $\{x \mid \f\}$ and a two-valued structure
\struct and returns the set $\{x \mid \f\}^\struct$. In the \idp language, this inference is invoked as \il{query(Q,S)}, where \il{Q} is a query and  \il{S} a structure. 
Continuing our
graph example,

\begin{lstlisting}
query Q:V{
    {x: (?y: Edge(x,y))}
}

procedure main(){
   models =modelexpand(T,S) 
   print(query(Q,models[1]))
}  
\end{lstlisting}
will print the set of all nodes that participate as the first node of an
edge in the first model computed by the model expansion. Note that
\il{query(...)} does not return a structure but a set.

\textbf{Model checking} and \textbf{satisfiability checking} are special
cases of model expansion.  In the former, the input structure is two-valued; the result of this inference is \emph{true} respectively \emph{false} if the input structure is a model of the theory. The latter also outputs a Boolean value, \emph{true} if the (possibly three-valued) output structure \emph{can be expanded} to a model. 
In the \idp language, both of these inference methods are called using \il{sat(T,S)}, where \il{T} is a theory and \il{S} a structure.\footnote{Note that satisfiability checking reduces to model checking in case \il{S} is two-valued.}

\textbf{Propagation} inference takes a theory and a structure and
returns a more precise structure that preserves all solutions. The
system supports different versions of propagation with different
costs. The most precise and most expensive version returns the partial
structure in which atoms are unknown iff they do not have the same
truth value in all models. This most expensive propagation is called using \il{optimalpropagate(T,S)}. 
Cheaper, approximate forms of propagation are called using \il{propagate(T,S)} and \il{groundpropagate(T,S)}.

\textbf{Deduction} takes as input an \foidaggpft theory \theory and an
FO theory $\theory_{FO}$ and returns true if $\theory\models
\theory_{FO}$, that is if the first theory logically entails the
second one. It is implemented in a sound but incomplete way by
translating \theory into a (weaker) \FO theory and calling the theorem
prover SPASS~\cite{cade/WeidenbachDFKSW09}. It is used internally in
the \idp system to detect and exploit functional dependencies in
predicates \cite{iclp/DeCatB13}. It is called using \il{entails(T1,T2)}.

\textbf{Symmetry detection} takes as input a theory \theory and a
partial structure \I and returns \emph{symmetries} over \theory and
\I. A symmetry is a function, say $f$, mapping structures to
structures, such that, for any two-valued expansion $J$ of \I that is
a model of \theory, $f(J)$ is also a model
~\cite{ictai/DevriendtBMDD12}. Symmetry detection also returns clauses
to break these symmetries and to eliminate symmetric models. 
Symmetry detection is not available as a Lua procedure but can be exploited in the model expansion workflow using the option \il{symmetrybreaking} (see Section \ref{sec:mx}).

\textbf{\D-model expansion} takes as input a definition \D and a
structure $\I_{in}$, interpreting all parameters of \D, and returns
the unique model \I that expands $\I_{in}$. This task is an instance
of model expansion, but is solved in \idp using different
technology. The close relationship between definitions and logic
programs under the well-founded semantics is exploited to translate \D
and $\I_{in}$ into a tabled Prolog program, after which XSB is used to
compute \I.  Taking an extra formula \f as input, with free variables
\xxx, the same approach is used to solve the query \f with respect to
\D and $\I_{in}$ in a goal-oriented way~\cite{tplp/Jansen13}. There is no dedicated Lua procedure for calling $\D$-model expansion. However, as described in Section \ref{sec:mx}, it is automatically detected that $\D$-model expansion can be performed in normal model expansion calls. 

\textbf{Unsat-core extraction} takes as input a theory \theory and a structure \I such that \theory has no models expanding \I. It returns a (minimal) theory $\theory_{out}$ entailed by \theory (obtained by instantiating some variables) such that $\theory_{out}$ still has no models expanding \I. Currently there is only support for printing the output theory, not for actually obtaining it; this procedures also prints from which line every sentence in $\theory_{out}$ was instantiated. This inference is particularly useful when debugging logical specification and can be called using \il{printunsatcore(T,S)}.

Finally, there is support for the \emph{linear time calculus} (LTC) defined by Bogaerts et al.~\cite{iclp/Bogaerts14}. One
can build an \il{LTCvocabulary} component as a vocabulary extending a
default LTC vocabulary and can use special inference methods to
initialize the state and to perform progression inference, i.e., to
infer the successor states step by step \cite{iclp/Bogaerts14}.

%% file: 4-mx.tex
\section{Under the Hood}\label{sec:mx}

In this section we focus on the inner working of some components of the IDP system. First, we discuss the workflow of the optimization inference and how users can control the various parts of the optimization engine. Afterwards we discuss techniques (under development) that help \idp scale to larger, possibly infinite, domains.

\newcommand{\optinference}{\mungrouped{OPT\langle V,\allowbreak \theory,\allowbreak \I,\allowbreak t,\allowbreak V_{out}\rangle}}

Optimization is the task of given a theory \theory, a structure \I, and a term $t$, all over the same vocabulary $V$, finding models of \theory that expand (are more precise than) \I.
This inference captures Herbrand model generation and (bounded) model expansion, both of which were proposed as logic-based methods for constraint solving, respectively in~\cite{tocl/EastT06} and~\cite{MitchellT05}.
In its most general form, we define optimization for typed \foidaggpf as follows.
The inference \optinference takes as input a theory $\theory$, structure $\I$ and term $t$, all over vocabulary $V$, and a vocabulary $V_{out}\subseteq V$.
Both \theory and $\I$ are well-typed and \I interprets all types.
The inference returns $V_{out}$-structures \J such that at least one model of \theory expanding both \I and \J exists and that expansion is minimal with respect to $t$.
The optimization inference is a generalization of the model expansion inference that takes the same arguments without the optimization term $t$ and that returns $V_{out}$-structures \J such that at least one model of \theory expanding both \I and \J exists. The workflow of these two inference methods coincides; for optimization, more search is needed to find \emph{optimal} models. 

One approach to optimization, used in \idp, is through \emph{ground-and-solve}: ground the input theory and term over the input structure and afterwards apply a search algorithm that, e.g., uses branch-and-bound to find optimal models.

In the rest of the section, we present how the optimization algorithm in \idp solves an \optinference task.
The workflow consists of an \foidaggpft grounding algorithm, a search algorithm for the full ground fragment of \foidaggpft and various analysis methods and transformations, that result in a smaller grounding and/or improved search performance.
%
%

The workflow of the optimization inference consists of three parts. First, theory and structure are preprocessed to optimize performance. Secondly, the theory is grounded into ECNF, the language supported by \minisatid. Last, the solver \minisatid is called to perform the actual inference on the ground theory.

\subsection{Preprocessing}\label{sec:preprocessing}
Several preprocessing steps are performed before the grounding phase. We briefly discuss them below.

\subsubsection{Checking structure consistency}

First, we specify the structure $\I$ that is parsed from a
structure component, in particular, how the type autocompletion works. The
following definitions formalize this.  The value of all other symbols 
is clear. 

We say that a type $t$ of a vocabulary $V$ is explicitly defined in structure
$\I$ if $\I$ contains an equation $t=S$.  We define the value of an
explicitly defined type in $\I$ as stated in its explicit definition.
For a type $t$ that is not explicitly defined, the interpretation of $t$ in $\I$ consists of all elements of its subtypes and domain elements that occur in the interpretation of symbols $\sigma$  at an argument position of type $t$. Formally: 
\[t^\I = \left(\bigcup_{\{s\mid \text{$s$ is a subtype of $t$}\}}s^\I\right)  \bigcup 
\left(\bigcup_{\{\sigma\in V\mid \text{the $i$'th type of $\sigma$ is $t$}\}}\{d_i\mid \ddd\in \sigma^ \I\}\right).\]

\ignore{Its explicit definition is a
definition consisting of rules $t(n).$ for each $n\in S$. (note, it is
a definition in the vocabulary with one unary predicate per type and
constant symbols per domain value identifier).

The type  autocompletion definition of a vocabulary $V$ for structure $\I$ is the definition defining all types of $V$, and consisting of, for each explicitly defined type, its explicit definition and for each other type $t$ with direct subtype $s$, the rule $\forall x (t(x)\rul s(x))$. The type completion of $V$ in $\I$  is the model of the type autocompletion definition in the Herbrand structure of the domain value identifiers. 
}
 
A number of constraints are imposed on structures:
\begin{itemize}
\item  
For any subtype $s$ of $t$, the interpretation of $s$ must be a subset of the interpretation of $t$:
  \[ \forall x:  s(x)\limplies t(x) \]
\item If $f$ is a partial or total function that is totally defined, there is at most one image for each value of the input: 
  \[  \forall \xxx, y, z: f(\xxx)=y \land f(\xxx)=z \limplies y=z\]
\item If $f$ is a totally defined total function, the structure must contain an image for each input argument:
  \[  \forall \xxx: \exists y: f(\xxx)=y.\]
\item A partially defined predicate cannot be both certainly true
  and certainly false for the same tuple:
  \[ \forall \xxx: \neg P\langle ct\rangle (\xxx)\lor \neg P\langle cf\rangle (\xxx)\]
\item For any partially defined function $f$, 
\begin{itemize}\item there is at most one certainly true image for every input:
  \[ \forall \xxx, y, z: f\langle ct\rangle (\xxx,y)\land f\langle
  ct\rangle (\xxx,z)\limplies y = z\]
\item if $f$ is a total function, at least one output is possible (not certainly false) for each input: 
  \[ \forall \xxx: \exists y: \lnot f\langle cf\rangle (\xxx,y)\]
\end{itemize}
\end{itemize}

When any of these constraints are violated in the autocompleted structure $\I$,
an appropriate error message is given. 

\subsubsection{Exploiting \inputstar-definitions}\cite{tplp/Jansen13}.

Assume that after preprocessing, we obtained a theory $\theory$ and a structure $\I_1=\I$. The next step is to eliminate  some  definitions of $\theory$ and extend $\I_1$.  The definitions that can be eliminated are the so called  \emph{\inputstar-definitions} of $\theory$ ~\cite{tplp/Jansen13}. 

We define inductively that a definition $\D$ of \theory is an  \inputstar-definition of  $\theory$ in structure $\I$ if all parameters of $\D$ have a 2-valued interpretation in $\I$ or are defined in \inputstar-definitions of \theory in $\I$. 

All \inputstar-definitions of $\theory$ can be evaluated in
advance. Essentially this is done by iterated $\D$-model expansion
steps: at each iteration an \inputstar-definition $\D$ is
selected that has all its parameters interpreted in the current
structure; we compute\footnote{This is done by translating it to a logic program and using \xsb.} its model in $\I_1$ and add
the interpretation of the defined symbols to $\I_1$. The
advantage of this is that as explained below, top-down grounding
techniques, as used in \idp, tend to be rather inefficient in case of
complex (inductive) definitions~\cite{phd/Wittocx10}. By evaluating
these definitions, we avoid grounding them and we make the
input structure more precise. 

Notice that this may result in inconsistency if the same predicate is
defined in multiple \inputstar-definitions that do not agree on
its value. Otherwise, the result is a theory $\theory_2$ and a refined
structure $\I_2$.

To exploit the above procedure for a maximal effect, it is extended
with a preprocessing step to split each definition $\D$ of $\theory$
in subdefinitions $\D_1,\dots,\D_n$. The advantage is that some
of these components may turn out to be \inputstar-definitions
whereas $\D$ is not. In this case we can evaluate part of $\D$.

The split of a definition is computed from its dependency relation.
Formally, the dependency relation $\leq$ of a definition $\D$ is the
least transitive relation containing all pairs $P\leq Q$ such that
defined symbol $P$ occurs in a rule defining $Q$. We say that $P \sim
Q$ if $P\leq Q\leq P$. This is an equivalence relation. The split of
$\D$ is the partition $\D_1,\dots,\D_n$ of \D such that each $\D_i$
defines an equivalence class of $\sim$. The idea is that each $\D_i$
defines a group of predicates that depend on each other.


\subsubsection{Delaying \outputstar-definitions}\cite{tplp/Jansen13,lash/BogaertsJDJBD14}
Consider a total definition $\D \in \theory_2$ such that the defined symbols of \D occur only in \D and are not interpreted in $\I_{2}$. 
Any structure that satisfies $\theory_2\setminus \{\D\}$ and does not interpret symbols defined in \D, can be extended to a model of $\theory_2$ by evaluating \D.
Consequently, there is no need to consider such a \D during search; we prefer to delay evaluation of \D as long as possible, to a postprocessing step. 

Such a \D is one example of an  \outputstar-definition~\cite{tplp/Jansen13,lash/BogaertsJDJBD14}. In general,  we define inductively that a definition $\D$ of \theory is an  \emph{\outputstar-definition} of  $\theory$ in structure $\I$ if all defined symbols of $\D$ only occur in $\D$ and in the bodies of rules of \outputstar-definitions.

These \outputstar-definitions do not have to be considered during search and can be evaluated afterwards in a post-processing step.
Theory $\theory_{3}$ is the theory obtained from $\theory_{2}$ by removing all \outputstar-definitions; this phase does not modify the structure, hence $\I_3=\I_2$.

\subsubsection{Reducing quantification depth using functional dependencies}\mycite{functionDetection}.
The size of the grounding is in general exponential in the nesting depth of quantifiers (as it involves the Cartesian product of the involved domain sizes).
One way to reduce the quantification depth, is to detect that symbols can be split into a number of symbols with a smaller arity.
Assume, for example, that a predicate \il{timeOf(session,time)} specifies at which time a certain session takes place in a scheduling application. 
If one could detect that the second argument \emph{functionally depends} on the first argument (the first uniquely determines the value of the second), then it could be replaced by a new function \il{timeFunc(session): time} instead.  With appropriate transformations, a subformula \il{?t: timeOf(s1,t) & timeOf(s2,t)} can then be reduced to \il{timeFunc(s1)=timeFunc(s2)}, eliminating the quantification over \il{t}.
Detection of functional dependencies is done using the deduction inference: we check whether an FO formula that expresses the dependency is entailed by the original theory.
For \il{timeOf}, the functional dependency holds if the theory entails the sentence $\forall s: \exists_{=1} t: \mathit{timeOf}(s,t)$.

This preprocessing phase takes as input the theory $\theory_{3}$ and structure $\I_{3}$ and returns $\theory_{4}$ and $\I_{4}$, in which entailed functional dependencies have been made explicit and quantifications have been dropped where possible.

However, as the user expects models in the original vocabulary, 
additional \outputstar-definitions are added to $\theory_4$, that define the original symbol in terms of the newly introduced ones.
In our example, this would be the definition 
\begin{lstlisting}
define {
  ! s t: timeOf(s,t) <- t = timeFunc(s).
}
\end{lstlisting}


\subsubsection{Exploiting symmetries}\label{sec:symmdetect}\cite{ictai/DevriendtBMDD12,sat/DevriendtBBD16}.
It is well-known that if a problem exhibits symmetries, they can cause a search algorithm to solve the same (sub)problem over and over again.
For example the ``pigeonhole'' problem ``do $n$ pigeons fit in $n-1$ holes?'' is known to be hard for SAT-solvers.
Symmetries can be detected and broken on the propositional level~\cite{shatter,sat/DevriendtBBD16}, but for large problems, even the task of detecting symmetries becomes infeasible.
Detecting symmetries on the first-order theory \cite{devriendt2016} is often an easier problem, as much more structure of the problem is explicitly available. For example for an \FO specification of the pigeonhole problem, it is almost trivial to detect that all pigeons are interchangeable.
The symmetry detection inference in \idp detects a simple, frequently occurring form of symmetries: locally interchangeable domain elements (see \cite{devriendt2016}). 
Two domain elements are considered interchangeable if they are of the same type and occur only symmetrically in interpreted predicates. 
Detected symmetries are handled by adding sentences to $\theory_4$  that statically break those symmetries, resulting in the theory $\theory_5$. 




\subsection{Ground-And-Solve}\label{ssec:ground-and-solve}

\subsubsection{Ground}\cite{ictai/DeCat13}.
The grounding algorithm visits the resulting theory ($\theory_{5}$) in a depth-first, top-down fashion, basically replacing all variables by all their matching instantiations, according to the interpretation of their types in a partial structure $\I_5$. For example, a formula $\forall x[t]: \psi(x)$ is replaced by $\bigand_{d\in t^{\I_5}} \psi(d)$. 

However, such an instantiation might be unnecessary large.
Indeed, if the value of a term or formula is known in the current structure $\I_5$ for a given instantiation of its free variables, it should not have been grounded in the first place. The solution is to reuse query inference.
For example, consider a formula $\forall \xxx [\TTT]: \f$ and an instantiation \ddd for the free variables \yyy of this formula.
In that case, \xxx need only be instantiated with tuples $\ddd'$ for which $\f[\yyy\subs\ddd,\xxx\subs\ddd']$ is not certainly true in $\I_5$. Finding such tuples can be done using the query inference on a derived structure over a vocabulary in which \il{<ct>} and \il{<cf>} tables have an explicit representation. 
For all instantiations of \xxx not in the result of that query, we are certain that the subformula is true anyway.
In fact, an incomplete (cheaper) query inference can be applied, as any over-approximation will result in additional grounding, still maintaining correctness.
The result is a ground \foidaggpft theory. Several optimizations for this step exist, as discussed in \cite{ppdp/JansenDDJ14}; below we discuss some of them.

\subsubsection{Simplification}
To ensure models are generated that expand the input structure, not only the ground theory, but also the structure $\I_5$ is passed to the search algorithm.
Since we use a top-down grounding algorithm, we can optimize over this:
whenever a domain atom or term is generated by instantiating variables, instead of using the atom or term itself, its interpretation is filled in in the grounding.
For example, if a formula $\forall x[t]: P(x)\lor Q(x)$ is grounded in a structure with interpretations 
\begin{lstlisting}
T={1;2;3} 
P<ct> = {1} 
Q<cf> = {3} 
\end{lstlisting}
a simplified grounding is 
\[(\ltrue \lor Q(1)) \land (P(2)\lor Q(2)) \land (P(3)\lor \lfalse).\]
This sentence can be simplified even more, by propagating derived truth values upwards, resulting in 
\[\ltrue \land (P(2)\lor Q(2)) \land P(3)\]
and finally
\[ (P(2)\lor Q(2)) \land P(3).\]

These simplification techniques can have as effect that large parts of the theory do not need to be grounded. For example, consider a sentence 
 $\forall x[t]: P(x)\lor \varphi(x)$, where $\varphi$ might be a large formula. For all instantiations $d$ of $x$ for which $P(d)$ holds, the formula $P(d)\lor \varphi(d)$ simplifies to $\ltrue$. Hence $\varphi(d)$ does not need to be grounded for such $d$. 




\subsubsection{Approximation and lifted unit propagation}\cite{\refto{GroundingWithBounds},acm/wittocx}.

The above grounding algorithm exploits information in the input structure $\I$ using the query inference. Essentially, it grounds only the instances $\varphi[\ddd]$ of formulas that are unknown in $\I$. As a consequence, more precise input structures $\I$ yield smaller groundings and increased search performance. 
This observation gave rise to the algorithms presented in~\mycite{GroundingWithBounds}, where instead of using structure $\I_5$ directly, we first compute a more precise structure $\I_6$  that approximates  all models of $\theory_5$ that expand  $\I_5$.  %
Ideally, we would like to compute  the \emph{most precise} structure that is less precise than all models of $\theory_5$ that expand $\I_5$.
Of course, finding this ideal structure is a task that is even harder than the original problem.

Instead of searching for this ideal structure, \idp's approach is to execute a lifted (approximative) version of the unit propagation that would occur after grounding.
The result is stored as a symbolic representation of a structure.
Namely, with each symbol $P$, we associate  two symbolic set expressions $S_{ct}$ and $S_{cf}$ with intended meaning that in structure $\I_6$, $P_{ct}$, respectively $P_{cf}$, is interpreted as  $S_{ct}^{\I_5}$, respectively $S_{cf}^{\I_5}$.   
Consider, for example, the following theory 
\begin{align*}
 &\forall x:P(x)\limplies Q(x).\\
 &\forall x: \lnot Q(x) \limplies R(x).
\end{align*}
Symbolic unit propagation then results in, e.g., a symbolic representation of $\I_6$ that interprets $Q_{ct}$ as $\{x \mid P_{ct}(x)\lor R_{cf}(x)\}$ (the latter interpreted in $\I_5$!), $R_{ct}$ as $\{x \mid Q_{cf}(x)\}$.
During the grounding phase, all queries for variable instantiations and the interpretation of atoms and terms are evaluated relative to this symbolic interpretation, resulting in fewer instantiations and more precise interpretations.
E.g., if $P$ is interpreted in $\I_5$, and $Q$ and $R$ are completely unknown in $\I_5$, then the second
sentence will only be instantiated for $x$'s such that that $P(x)$ is not true
in $\I_5$.

A symbolic representation of complete lifted unit presentation often consists of complex formulas, which are infeasible to query.
However, any approximation of those formulas is sufficient, as long as the resulting structure is at least as precise as $\I_5$.
Consequently, the formulas are simplified to balance the estimated cost of querying against the expected reduction in number of answers.

\subsubsection{Search}\label{sec:search}
Optimization in \idp relies on the search algorithm \minisatid~\cite{ictai/DeCat13} for ground \foidaggpft theories.
It takes the ground theory as input together with structure $\I_{5}$.
The algorithm combines techniques from SAT, \CP and \ASP through a DPLL(T) architecture~\mycite{DPLLT}.
At the core lies the SAT-solver \minisat~\cite{sat/EenS03}, a complete, Boolean search algorithm for propositional clauses.
This core is complemented by a range of ``propagator'' modules that take care of propagation for all other types of constraints in the theory, such as aggregates, definitions and atoms containing functions.
Each module is responsible for explaining its propagations in terms of the current assignment.
For a definition \D, for example, the module checks whether the current assignment satisfies \D's completion, whether the current assignment contains unfounded sets, and when a complete assignment is found, whether the structure is the well-founded model of \D.
Optimization is taken care of by a module that ensures the search space is visited in a branch-and-bound fashion.
Whenever a model $M$ is found, with value $v$ the interpretation of $c$ in $M$, a constraint $c<v$ is added to the ground theory (which raises a conflict, leading to backtracking and additional search).

\paragraph{On the importance of CP integration} \cite{ictai/DeCat13,ALP/DeCatBD14}
In contrast to previous versions, the current version of \minisatid supports ground \foidaggpft \emph{with function symbols}. Function symbols are handled using techniques from constraint programming \cite{cp/FeydyS09}. To illustrate the importance of having uninterpreted function symbols, consider the following birthday riddle.

\begin{example}
``To determine my age, it suffices to know that my age in 2013 is halfway between two consecutive primes, that my age's prime factors do not sum to a prime number, and that I was born in a prime year.''. In the \idp language, it can be modeled as:
\begin{lstlisting}
vocabulary V {
    type Nb isa nat
    Age: Nb         //uninterpreted constant: my age
    Prime(Nb)       //predicate containing all prime numbers
    YearOfBirth: Nb //uninterpreted constant: my year of birth
}
theory T : V  {
    //Definition of prime numbers
    define {
        ! x[Nb]: Prime(x) <-
            x>1 & 
            !y [Nb]: 1 < y < x =>  (x % y ~= 0). 
    }
    
    //Relation between age (in 2013) and year of birth
    Age = 2013 - YearOfBirth.
    
    //My age in 2013 is halfway between two consecutive primes
    ?x1 x2: 
        //x1 and x2 are prime
        Prime(x1) & Prime(x2) & 
        //they are consecutive
        ~(?y: Prime(y) & x1 < y < x2) & x1 < x2 &
        // my age is halfway between them
        Age = (x2 + x1)/2.
    
    //I was born in a prime year
    Prime(YearOfBirth).

    //my age's prime factors do not sum to a prime number
    ~Prime(sum{x: Prime(x) & 1 < x =< Age & Age % x = 0 : x}).
}

structure S : V { 
    Nb = {0..2013} 
}
\end{lstlisting}
\idp is unable to ground this theory without using uninterpreted function symbols (this is controlled through the option \il{cpsupport}, discussed below) due to memory exhaustion. With uninterpreted constants, \idp takes half a second to find a solution. In fact, \idp proves that 48 different solutions exist; however only one is an age below 100, namely $Age=26$.
\end{example}



\subsection{Post-Processing}\label{sec:postprocess}
As a final, post-processing step, structures returned by the search step are translated back to structures over $V$.
Next, they are merged with \I, \outputstar-definitions are evaluated over them and finally, they are projected to $V_{out}$, resulting in structures $\I_{out}$ that are solutions to the original  \optinference problem.

An \outputstar-definition is only evaluated if evaluating it will have an effect on the eventual $V_{out}$-structure.
If symmetry-breaking was applied, additional solutions can be generated by applying the symmetries to the solutions found.

\subsection{Controlling the optimization workflow}\label{sec:control}
Various components of the optimization workflow can be controlled using options. We provide a brief overview. 
\begin{description}
 \item[stdoptions.assumeconsistentinput]\textit{default:false}\hfill\\ If this option is true, the systems assumes that the input provided by the user is consistent and the consistency checks are skipped. Use at your own risk. 
 \item[stdoptions.xsb]\textit{default:true}\hfill\\ If this option is true, \inputstar-definitions are evaluated using the XSB Prolog system. Otherwise, they are evaluated using standard ground-and-solve techniques. We recommend to use XSB for efficiency reasons. 
 \item[stdoptions.postprocessdefs]\textit{default:true}\hfill\\ This option controls whether  \outputstar-definitions are delayed until after search. In general, we recommend to turn this option on. However, since detection of \outputstar-definitions is implemented through a bootstrapping approach, enabling might cause a small delay of up to a second. Hence, if your goal is to solve a large number of very small problems, we recommend to turn this off.  
 \item[stdoptions.splitdefs]\textit{default:true}\hfill\\ This option controls whether  definitions are split into minimal strata. As with the previous, we recommend to turn it on in most cases, unless when solving very small problems. 
 \item[stdoptions.symmetrybreaking]\textit{default:``none''}\hfill\\
 This option controls whether symmetries are broken and if they are broken, using which method. Currently, \idp only provides support for breaking symmetries \emph{statically}, hence this option can be set either to \il{``none''} or to \il{``static''}. Future versions of \idp might also offer the choice \il{``dynamic''}. Whether symmetry breaking is beneficial strongly depends on the problem at hand. 
 \item[stdoptions.reducedgrounding]\textit{default:true}\hfill\\ This option controls whether the grounding is simplified using information from the structure. We recommend to enable this option in most use cases. 
 \item[stdoptions.groundwithbounds]\textit{default:true}\hfill\\ This option controls whether the grounding size is reduced using the approximation techniques described above. We recommend enabling this option in most use cases. 
 \item[stdoptions.liftedunitpropagation]\textit{default:true}\hfill\\ This option controls whether the symbolic representation of the input structure ($\I_6$) is evaluated in advance, resulting in a concrete representation of  the input structure. We recommend enabling this option in most use cases. 
 \item[stdoptions.cpsupport]\textit{default:true}\hfill\\ This option controls whether function symbols are allowed in the grounding. If turned off, the ground theory will be entirely propositional, if turned on, functions symbols can appear in the ground theory; they are then handled by constraint programming techniques. We recommend enabling this option in most use cases, except for hard combinatorial problems with a very small grounding.
  \item[stdoptions.cpgroundatoms]\textit{default:false}\hfill\\ This option controls whether function symbols are allowed to occur \emph{nested} in the grounding. This is an advanced feature, with as advantage a smaller grounding, but as disadvantage possible loss of propagation.  Whether enabling this option is beneficial strongly depends on the problem at hand. 
 \item[stdoptions.functiondetection]\textit{default:false}\hfill\\ This option controls whether predicate symbols are automatically replaced by function symbols. Whether enabling this option is beneficial depends strongly on the problem at hand. For theories crafted by experts (and manually optimised), this options is probably not beneficial. The more naive the theory is, the more potential benefit this option has. If this option is enabled, we recommend to also enable the cpsupport option. 
 \item[stdoptions.nbmodels]\textit{default:1}\hfill\\ This option specifies the number of models that need to be returned by the optimization (or model expansion) inference. 
 \item[stdoptions.verbosity]\textit{default:0 (for all suboptions)}\hfill\\ Suboptions of this option control the verbosity of various components in the workflow. This is used mainly for debugging purposes, e.g., to know which part of the solver is causing certain delays. 
\end{description}


\subsection{Scalability and Infinity}\label{ssec:infinity}
The reader might have noticed that structures and groundings can be very large or even infinite (for example, when a predicate or a quantified variable is typed over \il{int}).
Model expansion (and thus, optimization) over infinite structures takes infinite time in general.
In \idp, several techniques are applied to address this issue; they often work well in practice.

A first such technique has been explained in Subsection \ref{ssec:ground-and-solve}. 
By intelligent reasoning over the entire theory, we can sometimes derive better variable bounds. 
Suppose for example that a theory contains formulas $\forall x[int]: P(x)\limplies Q(x)$ and $\forall y[int]: P(y) \limplies R(y)$, where $Q$ only ranges over a finite type, say $T$, but $P$ and $R$ range over $int$.
The first of these sentences guarantees that $P$ will only hold for values such that $Q$ holds, hence $P$ can only hold for values in the finite type $T$. 
Thus we know that the second sentence should only be instantiated for $y$'s in $T$, i.e., by deriving an improved bound for $y$, the grounding of the second sentence suddenly becomes finite. The first sentence can be handled similarly. We only ground this sentence for $y$'s in $T$ and maintain a symbolic interpretation expressing that $P$ is certainly false outside of $T$.

Second, the usage of a top-down, depth-first grounding algorithm has the advantage that interpretations can be evaluated \emph{lazily}:
(i) instantiations of quantifications can be generated one at a time, and  
(ii) the interpretation of atoms and terms needs only to be retrieved for atoms and terms that effectively occur in the grounding.
The same advantage applies for symbols that are interpreted by (complex) procedures: the procedures are only executed for relevant occurrences of that symbol.

Third, the search algorithm maintains bounds on the interpretation of function terms, taking constraints in the grounding into account.
Consider a constant $c: int$, which in itself would result in an infinite search space.
However, combined with, e.g., a constraint $0\leq c \leq 10$ in the grounding, the solver reduces $c : int$ to $c : [0,10]$, a finite search space.

A fourth technique currently under development to increase scalability is \emph{Lazy Grounding} \cite{jair/CatDBS15}. 
Lazy Grounding is based on the observation that the entire grounding often is not necessary to find solutions to a model expansion or optimisation problem.
Instead, the technique interleaves grounding with search as follows.
Initially, it (roughly) splits the input theory into two parts: one part is grounded and the underlying solver performs its standard search algorithm on it.
The other part of the theory is \emph{delayed}: the system makes assumptions about it that guarantee that models found by the solver can be extended to models of the entire theory. 
Whenever these assumptions become violated, i.e., when they are inconsistent with the solver's current assignment, the splitting of the theory is revised.
Consider for example $\forall x[int]: P(x) \limplies
\f\label{f:first}$ with $\f$ a possibly large formula; it has an
infinite grounding.
A smart lazy grounder could delay the grounding of that sentence with the assumption that $P$ is false for all integers.
During search, only when an atom $P(d)$ becomes true, is the sentence grounded for $x=d$ and only that ground sentence is added to the search, the remainder still delayed on the assumption that $P$ is false.
Whenever the search algorithm finds a structure that satisfies the grounding and does not violate any assumptions, that structure can be straightforwardly extended to a model of the whole theory.

%% file: 5-modeling.tex
\section{In Practice}\label{sec:practice}

Both \idp and its search algorithm \minisatid are open-source systems, freely available from \url{dtai.cs.kuleuven.be/krr/software}.
Next to accepting input in the \idp language, both systems provide \texttt{C++} interfaces.
The search algorithm \minisatid supports input in Clausal Normal Form (CNF), Quantified Boolean Form (QBF, CNF's higher-order relative) \cite{aaai/EglyETW00}, ground ASP (in the LParse-Smodels intermediate format~\cite{Syrjanen98}) and FlatZinc~\mycite{MiniZinc}.

\idp can be be tried online in our web-IDE at \url{https://dtai.cs.kuleuven.be/krr/idp-ide/}. Several modelling examples are available in the web-IDE as well as at \url{https://dtai.cs.kuleuven.be/software/idp/examples}. 
The web-IDE also provides support for the visualisations of structures, as explained, e.g.,  in \url{https://dtai.cs.kuleuven.be/krr/idp-ide/?chapter=intro/9-IDPD3}. 

The main usage of \idpthree is currently its model expansion inference, which as discussed earlier, is closely related to generating answer sets of logic programs and  to solving constraint satisfaction problems.
As such, it shares applications with those domains, general examples of which are scheduling, planning, verification and configuration problems.
More concretely, some applications have been modelled in~\cite{TPLP/BruynoogheBBDDJLRDV}, demonstrating its applicability as both an approach to replace procedural programming in some cases and as an approach to rapid prototyping due to the short development time.
It has been used to analyze security issues in several contexts, with an emphasis on formal approaches that allow intuitive modelling of the involved knowledge~\cite{essos/DecroixLDN13,phd/heyman}.
The model expansion engine, and various other types of inference, have been used for interactive configuration \cite{padl/HertumDJD16,tplp/HertumDJD17}. 

The performance of \idp has been demonstrated for example in the ASP competition series, in 2009~\mycite{ASPComp2} (\idptwo), 2011~\mycite{ASPComp3} (\idptwo) and 2013~\mycite{ASPComp4} (\idpthree) and in~\cite{TPLP/BruynoogheBBDDJLRDV}, where it is compared to various existing approaches to specific problems.
The performance of the search algorithm \minisatid has been demonstrated in~\cite{cpaior/AmadiniGM13,corr/AmadiniGM13}, where it turned out to be the single-best solver in their MiniZinc portfolio, and in the latest Minizinc challenges~\cite{url:MinizincChallenge2012}.
In~\cite{TPLP/BruynoogheBBDDJLRDV}, it is demonstrated that a KR system like \idp can be practically used as a front-end for lower-level solvers (e.g., SAT-solvers), instead of manually encoding problems in SAT or using custom scripting.  
Experimental results showed that \idp is able to relieve this burden with minimal performance loss and greatly reduced development effort.

\idpthree is used as a didactic tool in various logic-oriented courses at various universities.

%% file: 6-related.tex
\section{Related Work}\label{sec:related}
Within several domains, research is targeting expressive specification languages and (to a lesser extent) multiple inference techniques within one language.
While we do not aim at an extensive survey of related languages (e.g., \cite{constraints/MarriottNRSBW08} has a section with such a survey), we do compare with a couple of them.

The B language~\cite{Abrial96}, a successor of Z, is a formal specification language developed specifically for the generation of procedural code. It is based on first-order logic and set theory, and supports quantification over sets. Event-B is a variant for the specification of event-based applications.
The language Zinc, developed by Marriott et al.~\cite{constraints/MarriottNRSBW08}, is a successor of OPL and intended as a specification language for constraint programming applications (mainly CSP and COP solving). It is based on first-order logic, type theory and constraint programming languages.
Within \ASP, a number of related languages, originating from logic
programs, are being developed, such as Gringo~\cite{lpnmr/GebserKOST09} and DLV~\cite{tocl/LeonePFEGPS06}. They support definitional knowledge and default reasoning. Implementations exist for inference techniques like stable model generation (related to model expansion), visualisation, optimization and debugging.
A comparison of ASP and \foid can be found in~\mycite{fodot2asp}.
The language of the Alloy~\cite{tosem/Jackson02} system is basically first-order logic extended with relational algebra operators, but with an object-oriented syntax, making it more natural to express knowledge from application domains centered around agents and their roles, e.g., security analysis.

The following are alternative approaches to model expansion (or to closely related inference tasks).
The solver-independent \CP language Zinc~\cite{constraints/MarriottNRSBW08} is grounded to the language MiniZinc~\mycite{MiniZinc}, supported by a range of search algorithms using various paradigms, as can be seen on \url{www.minizinc.org/challenge2012/results2012.html}.
In the context of constraint ASP (CASP), several systems ground to ASP extended with constraint atoms, such as Clingcon~\cite{tplp/OstrowskiS12} and EZ(CSP)~\mycite{EZCSP}.
For search, Clingcon combines the ASP solver Clasp~\cite{ai/GebserKS12} with the CSP solver Gecode~\cite{gecode}, while EZ(CSP) combines an off-the-shelf ASP solver with an off-the-shelf CLP-Prolog system.
The prototype CASP solver Inca~\mycite{Inca} searches for answer sets of a ground CASP program by applying \LCG for arithmetic and all-different constraints.
As opposed to extending the search algorithm, a different approach is to transform a CASP program to a pure ASP program~\cite{ijcai/DrescherW11}, afterwards applying any off-the-shelf ASP solver.
CASP languages generally only allow a restricted set of expressions to occur in constraint atoms and impose conditions on where constraint atoms can occur.
For example, none of the languages allows general atoms $P(\ccc)$ with $P$ being an uninterpreted predicate symbol.
One exception is $\mathcal{AC(C)}$, a language aimed at integrating ASP and Constraint Logic Programming~\cite{aclanguage}.
As shown in~\cite{aaai/Lierler12}, the language captures the languages of both Clingcon and EZ(CSP); however, only subsets of the language are implemented~\cite{lash/Gelfond08}.
